\documentclass[sigconf]{acmart} 

\usepackage{amsmath}
\DeclareMathOperator{\sign}{sign}
\usepackage{multirow}
\usepackage{enumitem}
\setlist[itemize]{leftmargin=*}
\setlist[enumerate]{leftmargin=*}
\allowdisplaybreaks
\AtBeginDocument{%
  }

\copyrightyear{2024}
\acmYear{2024}
\setcopyright{acmlicensed}
\acmConference[ICAIF '24]{5th ACM International Conference on AI in Finance}{November 14--17, 2024}{Brooklyn, NY, USA}
\acmBooktitle{5th ACM International Conference on AI in Finance (ICAIF '24), November 14--17, 2024, Brooklyn, NY, USA}
\acmDOI{10.1145/3677052.3698599}
\acmISBN{979-8-4007-1081-0/24/11}

\begin{document}

\title{Tax Credits and Household Behavior: The Roles of Myopic Decision-Making and Liquidity in a Simulated Economy}

\author{Kshama Dwarakanath}
\affiliation{%
  \institution{JP Morgan AI Research}
  \city{San Francisco}
  \state{California}
  \country{USA}
}
\email{kshama.dwarakanath@jpmorgan.com}

\author{Jialin Dong}
\affiliation{%
  \institution{University of California, Los Angeles}
  \city{Los Angeles}
  \state{California}
  \country{USA}
}

\author{Svitlana Vyetrenko}
\affiliation{%
  \institution{JP Morgan AI Research}
  \city{San Francisco}
  \state{California}
  \country{USA}
}


\begin{abstract}

There has been a growing interest in multi-agent simulators in the domain of economic modeling. However, contemporary research often involves developing reinforcement learning (RL) based models that focus solely on a single type of agents, such as households, firms, or the government. Such an approach overlooks the adaptation of interacting agents thereby failing to capture the complexity of real-world economic systems. 
In this work, we consider a multi-agent simulator comprised of RL agents of numerous types, including heterogeneous households, firm, central bank and government. In particular, we focus on the crucial role of the government in distributing tax credits to households. 
We conduct two broad categories of comprehensive experiments dealing with the impact of tax credits on 1) households with varied degrees of myopia (short-sightedness in spending and saving decisions), and 2) households with diverse liquidity profiles. 
The first category of experiments examines the impact of the frequency of tax credits (e.g. annual vs quarterly) on consumption patterns of myopic households. The second category of experiments focuses on the impact of varying tax credit distribution strategies on households with differing liquidities. 
We validate our simulation model by reproducing trends observed in real households upon receipt of unforeseen, uniform tax credits, as documented in a JPMorgan Chase report.
Based on the results of the latter, we propose an innovative tax credit distribution strategy for the government to reduce inequality among households. We demonstrate the efficacy of this strategy in improving social welfare in our simulation results.
\end{abstract}

\maketitle

\section{Introduction}
The spotlight in economic modeling has increasingly turned towards agent-based simulators which are well-suited to model real-world complexity and agent heterogeneity \cite{deissenberg2008eurace, haldane2019drawing}. In simulated economies, reinforcement learning (RL) emerges as a popular tool for determining optimal agent strategies, as demonstrated in recent work \cite{axtell2022agent, liu2022welfare, hill2021solving, chen2021deep}. New studies such as \cite{mi2023taxai, zheng2022, curry2022analyzing, dwarakanath2024abides} have explored the interactions among households, firms, the central bank and the government in simulated economic systems with RL agents. Such simulation tools coupled with learning techniques provide a flexible platform to answer economic policy questions \cite{dosi2019more}.

Still, aforementioned research has two limitations. Firstly, studies such as \cite{axtell2022agent, liu2022welfare, hill2021solving, mi2023taxai} restrict the application of RL to a limited set of agents, thus failing to incorporate other adaptive agents like the government or the central bank. This restriction overlooks complexities of real-world economic systems. 
Secondly, state-of-the-art such as \cite{curry2022analyzing} apply uniform tax credit distribution and overlook the crucial role of government intervention in maintaining social welfare through tax collection and redistribution. Although a recent paper \cite{liu2022welfare} introduces an algorithm to achieve both social welfare maximization and competitive equilibrium, it only accounts for two types of economic agents: agents and planners. 

Here, we study the impact of tax credits on heterogeneous household behavior, so as to inform our design of a tax credit distribution strategy that improves social welfare. 
We model household heterogeneity as arising from two sources: myopic decision-making and liquidity.
A large body of literature provides evidence of myopic decision-making in humans \cite{strotz1973myopia,green1994discounting,laibson1997golden}. Notably, human households vary in their degrees of myopia i.e., how far into the future they look when making their consumption-savings decisions \cite{chabris2010intertemporal}. 
Yet, there is limited work on isolating the effects of myopia on household behavior upon receipt of tax credits, especially accounting for heterogeneity and long time horizons.
Similarly, an extensive report by JPMorgan Chase \cite{JPMC} shows that household liquidity, defined as the ratio of savings to typical spending, is a strong predictor of consumption behavior when tax credits are distributed. They find that low-liquidity households see bigger increases in consumption spending than high-liquidity households upon receipt of tax credits. 

Gaining insights into household spending patterns, influenced by differences in myopia or liquidity, can offer valuable guidance to economic policymakers in crafting effective tax credit distribution schemes.
In this work, we enable RL strategies for all agents thereby creating a more realistic representation of the economy. We conduct a thorough investigation of the impact of uniform tax credits and their frequency on diverse household behavior. Based on our findings, we design a government policy for credit distribution that promotes social welfare. Our contributions are summarized below. \begin{itemize}
\item We consider a multi-agent economic model comprising households, a firm, central bank, and government. Our experiments utilize the economic simulator ABIDES-Economist \cite{dwarakanath2024abides}, an extension of a state-of-the-art financial market simulator \cite{byrd2019abides} tailored to economic simulations. In particular, we focus on evaluating the impact of tax credits on heterogeneous households with different degrees of myopia and liquidity.

\item To study how myopia impacts household behavior with tax credits, we model myopia by varying the reinforcement learning discount factor, which controls the relative importance of immediate and long term rewards. We then compare annual and quarterly tax credit frequency regimes. Experiments demonstrate household tendency to temporarily reduce labor hours when they receive annual credits, compared to more consistent labor with quarterly credits. We also observe that savings decrease and consumption increases as household myopia increases. 

\item To study how liquidity impacts household behavior with tax credits, we model liquidity preferences using parameters of household utility from consumption and savings. We then evaluate the impact of unforeseen, uniform tax credits on high-, medium- and low-liquidity households. Experiments show that low-liquidity households experience a greater increase in consumption upon credit receipt compared to high-liquidity households. 
These findings align with a JPMorgan Chase report based on real consumer data \cite{JPMC}, and serve as a validation of our economic model. 

\item Based on insights on household inequality due to unequal spending patterns with uniform tax credit distribution, we propose a novel tax credit distribution strategy for the government. Simulation results demonstrate its efficacy in improving social welfare.
\end{itemize}

\section{Background and Related Work}
\paragraph{Agent-based modeling in Economics.}
The field of Agent-based Computational Economics (ACE) focuses on using agent-based models (ABMs) to simulate interactions between economic agents. ABMs account for complex factors like agent heterogeneity, adaptation and the ability to model dynamics out of equilibrium compared to Dynamic Stochastic General Equilibrium (DSGE) models\cite{tesfatsion2006handbook}. 
\cite{deissenberg2008eurace} aimed to construct a large-scale comprehensive ABM of the European economy. Initial simulations targeted labor market dynamics with capital and consumer good firms and households with rule-based strategies. This led to subsequent extensions that captured the interplay between labor markets, industry evolution, credit markets and consumption \cite{dawid2016heterogeneous}.
There is significant work on ABMs of endogenous growth and business cycles which are empirically validated by replicating a set of microeconomic and macroeconomic stylized facts \cite{dosi2006evolutionary,dosi2017micro}\footnote{See \cite{dawid2018agent} for a survey of work related to ABMs for macroeconomic analysis.}. They have rule-based agents with predefined behaviors, providing a framework to compare such behavioral rules, particularly in the context of monetary and fiscal policy \cite{dosi2015fiscal}.

\paragraph{Agent-based economic modeling and Reinforcement Learning (RL)}
Since economic models depict households as entities maximizing their discounted sum of utilities over time, RL techniques are particularly useful to model household behavior \cite{atashbar2023ai}. E.g., \cite{chen2021deep} determine optimal consumption, saving and working strategies for a representative household using RL in a DSGE model given in \cite{evans2005policy}. Also, \cite{hill2021solving} learn consumption and labor strategies for discrete, heterogeneous households using RL in macroeconomic models combined with epidemiological effects under equilibrium.
Other works have employed RL to determine optimal economic strategies, including \cite{koster2022human} for redistribution of shared revenue, \cite{chen2021deep,hinterlang2021optimal} for the central bank's monetary policy, and \cite{zheng2022} for the government's tax policy.

\paragraph{Agent-based economic modeling and Multi-agent Reinforcement Learning (MARL)}
\cite{zheng2022} pioneered MARL in economic ABMs by studying tax policy design with agents and a planner. 
The planner learns to set marginal tax rates for each income bracket to balance equality and productivity, while agents learn to maximize their endowment utility within a gather-and-build game. 
\cite{curry2022analyzing} present a macroeconomic real-business-cycle ABM with consumers, firms, and the government, using MARL for all agents' strategies. However, the model has several limitations: it assumes a complete and uniform redistribution of income taxes to households, omits inventory holding risk for firms, and does not account for the central bank's role in monetary policy.
\cite{mi2023taxai} develop an economic simulator focused on taxation problems using MARL, where households maximize isoelastic consumption utilities and the government seeks to improve social welfare and economic growth. Despite scaling to 10,000 household agents, they assume market clearing, fixed interest rates, and have a rule-based firm agent. 
\cite{brusatin2024simulating} use MARL within a macroeconomic ABM with capital and credit to learn price-quantity strategies for consumer-goods producing firms. This while households, capital-goods firms and banks follow fixed strategies.
While only a few recent works delve into MARL within agent-based economic models, the list is slowly growing \cite{mi2023taxai,dwarakanath2024abides}. But, none focus on the design and impact of tax credit policy on heterogeneous households, in presence of other RL economic agents.

\paragraph{Myopic decision-making in humans.}
Humans are known to be prefer immediate rewards over those got later in the future. This short-sighted behavior (or myopia) has historically been documented in numerous studies and is modeled by temporal discounting of rewards \cite{mazur1985probability,green1994discounting,chabris2010intertemporal}. Two widely used models for temporal discounting include the exponential model \cite{samuelson1937note} and the hyperbolic model \cite{ainslie1992picoeconomics}\footnote{For amenability to standard RL algorithms, we use the exponential model \cite{liu2022biased}.
}. The discount factor determines the extent of myopia in these models, and varies widely among individuals. 
E.g., \cite{green1994discounting,green1996temporal} seek to measure the impact of age and income on temporal discount factors of individuals. 
Notably, \cite{green1996temporal} observe that lower income older adults showed a greater degree of temporal discounting than did both upper income older adults and upper income younger adults. 

\paragraph{Myopia and household decision-making.}
Intertemporal preferences have major consequences in household decisions and policy questions \cite{ericson2019intertemporal}.
There is work on designing optimal retirement schemes in a society with both rational and myopic agents \cite{feldstein1985optimal,cremer2011myopia}. 
Also, \cite{kaplow2015myopia} studies the impact of taxation on labor supply decisions of myopic individuals.
These studies focus on short horizon problems (typically two period) using retirement plans to compel myopic agents to save. 
Here, we examine the effects of myopia on household consumption and labor decisions in longer horizon problems, with specific focus on the impact of tax credit frequency. 
Myopia inevitably affects household response to tax credits, particularly when they are distributed once a year versus regularly every quarter.

\paragraph{Tax Credit studies}\label{subsec:jpm_report}
We aim to study the impact of tax credits on household consumption and labor decisions. 
Focusing on the report in \cite{JPMC} which uses customer transaction data to estimate the impact of advanced Child Tax Credit payments on household spending\footnote{See \cite{JPMC} for a dataset description, including transactions for 2.4 million households whose members had active checking accounts between January 2019 and January 2022.}, the findings reveal two key insights. Firstly, spending spiked immediately after receiving credits \cite{JPMC_tax_time}. Secondly, household liquidity (defined as the amount of cash on hand divided by spending) was a stronger predictor of consumption response than income. Specifically, low-liquidity households exhibited larger increases in consumption spending after receiving credits compared to high-liquidity households.
A key limitation of this analysis is the assumption that spending behavior of non-recipients of tax credits is similar to that of recipients had they not received the credits. This assumption is used to isolate the effects of tax credits from household-level differences on their propensities to consume. 
Also, there are differing opinions on the impact of tax credits (particularly, child tax credits) on the labor force even using real data. E.g., \cite{ananat2022effects} predict minimal impact while \cite{NAS2019,corinth2021anti} foresee employment reductions with workers exiting the labor force. 
This is where simulation techniques prove useful as they allow for the isolation of policy change effects on the same households before and after implementation.

\section{Economic model}\label{sec:econ_model}
In our economic model, households earn labor income and consume goods from a firm, the firm produces goods using household labor \cite{krusell1998income}, the government collects income tax from households and distributes tax credits, and the central bank sets interest rate to achieve target inflation and improve gross domestic product (GDP) \cite{kaplan2018monetary}.
Unlike previous work, every agent in our system employs RL to learn a strategy that maximizes individual objectives. 
Our model with multiple RL agents is formalized as a Markov Game (MG), where each agent has partial observability of the global system state. 
The MG consists of finite-length episodes of $H$ time steps where each time step $t\in\{0,1,\cdots,H-1\}$ corresponds to one-quarter of economic simulation. Table \ref{table} contains a summary of agent observations, actions, parameters and rewards.
\subsection{Households}\label{subsec:econ_h}
	At time step $t$, household $i$ works for $n_{t,i}$ hours and requests to consume $c^{req}_{t,i}$ units of the good produced by the firm. 
	The good's price is set by the firm as $p_{t}$, and the income tax rate is set by the government as $\tau_t$. The realized household consumption depends on available inventory $Y_{t}$ and total demand $\sum_k c_{t,k}^{req}$. If there is insufficient supply to meet demand, goods are rationed proportionally as $c_{t, i}=\min \left\{c_{t,i}^{req}, Y_{t} \cdot \frac{c_{t,i}^{req}}{\sum_k c_{t, k}^{req}}\right\}$ so that household $i$ pays a cost of consumption of $c_{t,i}p_{t}$.
	Savings $m_{t,i}$ increase according to interest rate $r_t$ set by the central bank, while also being affected by the cost of consumption and labor income equal to $n_{t,i}w_{t}$ where the firm pays a wage of $w_{t}$. Moreover, household $i$ pays an income tax amounting to $\tau_tn_{t,i}w_{t}$, where the government redistributes a portion of the tax revenue as tax credits $\kappa_{t,i}$. Household savings evolve accordingly as $m_{t+1, i}=\left(1+r_t\right) m_{t, i}+\left(n_{t, i} w_{t}-c_{t, i}p_{t}\right)-{\tau_tn_{t,i}w_t}+\kappa_{t, i}$.
As in \cite{JPMC}, we define the liquidity of household $i$ at time step $t$ as\begin{align}
     l_{t,i}=\frac{m_{t,i}}{c_{t,i}p_t}\label{eq:h_liquidity}.
 \end{align}
Note that $l_{t,i}$ is a measure of the household's cash buffer measuring its savings relative to consumption spending, so that low-liquidity households are more cash-constrained than high-liquidity ones. 
 
	Each household $i$ optimizes its labor hours and consumption to maximize utility \cite{evans2005policy}:
 \begin{align}
     \max_{\lbrace n_{t, i}, c_{t, i}^{req}\rbrace_{t=0}^{H-1}} \sum_{t=0}^{H-1} \beta_{i, \mathbf{H}}^t  u\left(c_{t,i},n_{t,i}, m_{t+1,i};\gamma_i,\nu_i,\mu_i\right)\nonumber\\
     \textnormal{where }u(c, n, m ; \gamma, \nu, \mu)=\frac{c^{1-\gamma}}{1-\gamma}-\nu n^2+\mu\sign(m)\frac{|m|^{1-\gamma}}{1-\gamma} \label{reward_h}
	\end{align}
	where $\beta_{i, \mathbf{H}}\in[0,1]$ is the discount factor. A household with $\beta_{i, \mathbf{H}}$ close to $1$ discounts future rewards less than one with $\beta_{i, \mathbf{H}}$ close to $0$. So, less myopic (far-sighted) households have high $\beta_{i, \mathbf{H}}$ compared to more myopic (near-sighted) households with low $\beta_{i, \mathbf{H}}$. The utility function is a sum of isoelastic utility from consumption with parameter $\gamma_i$, isoelastic utility from savings with parameter $\gamma_i$ and coefficient $\mu_i$, and a quadratic disutility of work with coefficient $\nu_i$.
	
	\subsection{Firm}
	The firm employs households to produce goods and earns money from consumed goods. At time step $t$, it determines the good's price $p_{t+1}$ and wage $w_{t+1}$, which will come into effect at time step $t + 1$.
	Based on total labor hours $\sum_i n_{t,i}$, the firm produces $y_{t}$ units of goods per a Cobb-Douglas production function: 
	\begin{equation}\label{production}
	y_{t}=\epsilon_{t}\left(\sum_i n_{t, i}\right)^{\alpha}
	\end{equation}
	where $0 \leq \alpha \leq 1$ characterizes the efficiency of production using labor\footnote{(\ref{production}) is passed through a floor function to compute integer number of goods produced.}. And, $\epsilon_{t}$ is the exogenous production factor which follows a log-autoregressive process with coefficient $0 \leq \rho \leq 1$, given by $\epsilon_{t}=\left(\epsilon_{t-1}\right)^{\rho} \exp \left(\varepsilon_{t}\right)$ and allows for an exponential increase with a shock $\varepsilon_{t}\sim\mathcal{N}(0,\sigma^2)$. As households purchase $ \sum_i c_{t,i}$ units of goods, the inventory of the firm evolves as $Y_{t+1} = Y_{t} + y_{t} - \sum_i c_{t,i}$.
 
 The firm optimizes the price of its goods and wage of households to maximize profits as in the following objective:
 \begin{equation}\label{reward_F}
     \max_{\lbrace w_{t}, p_{t}\rbrace_{t=0}^{H-1}} \sum_{t=0}^{H-1} \beta_{ \mathbf{F}}^t\left(p_{t} \sum_ic_{t, i}-w_{t}\sum_in_{t, i}-\chi p_{t} Y_{t+1}\right)
 \end{equation}
	where $\beta_{ \mathbf{F}}\in[0,1]$ is the firm's discount factor and $\chi>0$ is the coefficient for the inventory risk term that prevents excessive price increases for profits without a subsequent increase in consumption. This term captures the risk of holding inventory due to producing much more than is consumed in absence of market clearing.
 
\subsection{Central Bank}
 \begin{table*}[t]
    \centering
    \caption{Summary of observations, actions, parameters and rewards of agents in our economic model.}
    \label{table}
    \begin{tabular}{llllc}
    \toprule
    \textbf{Agent} & \textbf{Observation} & \textbf{Action} & \textbf{Parameter} & \textbf{Reward} \\\midrule
    \multirow{8}{*}{Household $i$} & $(m_{t,i},r_{t},\tau_{t},\kappa_{t,i},p_{t},w_{t})$ & $(n_{t,i},c_{t,i}^{req})$ & $(\gamma_i,\nu_i,\mu_i)$ & (\ref{reward_h}) \\
    & $m_{t,i}$: saving & $n_{t,i}$: labor hours & $\gamma_i$: parameter for isoelastic utility&\\ 
    & $r_t$: interest rate & $c_{t,i}^{req}$: requested consumption &\hspace{1em} of consumption and savings&\\ 
    & $\tau_{t}$: tax rate &&$\nu_i$: {coefficient for labor disutility} &\\ 
    & $\kappa_{t,i}$: tax credit &&$\mu_i$: {coefficient for savings utility} &\\ 
    & $p_{t}$: price of good set by firm &&&\\ 
    & $w_{t}$: wage set by firm &&&\\\midrule
    \multirow{5}{*}{Firm} & $(\varepsilon_t,{\epsilon_{t-1}}, Y_{t},{p_{t}, }w_{t},\sum_i n_{t,i}, \sum_i c_{t,i})$ & $(w_{t+1},p_{t+1})$ & $(\alpha,\rho,\sigma,\chi)$ & (\ref{reward_F})\\
    & $\varepsilon_{t}$: production shock &&$\alpha$: efficiency of production using labor &\\ 
& $\epsilon_{t-1}$: exogenous production factor &&$\rho$: auto-regression coefficient for $\lbrace\epsilon_t:t\geq0\rbrace$&\\ 
    & $Y_{t}$: inventory of firm && $\sigma$: standard deviation of production shock&\\ 
    & $c_{t,i}$: realized consumption&&$\chi$: coefficient for inventory risk &\\\midrule
    \multirow{3}{*}{Central Bank} &$(p_{t-4}, p_{t-3}, p_{t-2}, p_{t-1}, p_t, y_{t})$&$r_{t+1}$ &$(\pi^\star,\lambda)$&(\ref{reward_cb})\\
    & $y_t$: production of firm &&$\pi^\star$: target inflation&\\
    &&&$\lambda$: coefficient for GDP objective\\\midrule
    \multirow{4}{*}{Government} &$(\tau_t,\{\kappa_{t,i}\}_i,\{l_{t,i}\}_i,\sum_i\tau_tn_{t,i}w_t,\mathcal{H}_t)$& $(\tau_{t+1},\{\kappa_{t+1,i}\}_i)$&$(\xi,\Delta T,\theta)$&(\ref{reward_g})\\
    &$l_{t,i}$: liquidity of household $i$ &&$\xi$: Fraction of taxes redistributed as credits&\\
    &$\mathcal{H}_t$: sum of household utilities&&$\Delta T$: Tax credit frequency in quarters &\\
    &&&$\theta$: Coefficient for household utility&\\\bottomrule
    \end{tabular}
\end{table*}
	The central bank gathers data on annual price changes and firm production at each time step. At time step $t$, the central bank sets the interest rate $r_{t+1}$, which will come into effect at time step $t+1$. The inflation rate at $t$ is derived by the central bank based on annual price changes as $\pi_t=\frac{ p_{t}}{ p_{t-4}}$, and the GDP is measured by firm production $ y_{t}$.
	The central bank optimizes monetary policy to achieve inflation and GDP targets as follows \cite{svensson2020monetary,hinterlang2021optimal}:
	\begin{align}\label{reward_cb}
	\max _{\lbrace r_t\rbrace_{t=0}^{H-1}} \sum_{t=0}^{H-1} \beta_{\bf{CB}}^t\left(-\left(\pi_t-\pi^{\star}\right)^2+\lambda y_{t}^2\right)
	\end{align}
	where $\beta_{\mathbf{CB}}\in[0,1]$ is the central bank's discount factor, $\pi^{\star}$ is the target inflation and $\lambda>0$ is the GDP weight relative to inflation. 
	
	\subsection{Government}\label{subsec:econ_gov}
   At time step $t$, the government collects income taxes amounting to $\sum_i\tau_tn_{t,i}w_t$, and sets tax rate $\tau_{t+1}$ that will come into effect at time step $t + 1$.
 It also distributes a portion $\xi\in[0,1]$ of the total taxes collected in a year as tax credits to households in the next year, given every $\Delta T$ quarters. 
 The total tax collected in year $\mathit{y}$ is given by $\mathcal{T}_{\mathit{y}}=\sum_{t=4\mathit{y}}^{4\mathit{y}+3}\sum_i\tau_tn_{t,i}w_t$ where the outer sum is over quarterly time steps in year $\mathit{y}$. 
 Let $f_{\mathit{y}+1,i}$ denote the fraction of annual credits distributed to household $i$ over the next year $\mathit{y}+1$ so that its total credit in year $\mathit{y}+1$ is $\kappa_{\mathit{y}+1,i}=f_{\mathit{y}+1,i}\xi\mathcal{T}_{\mathit{y}}$.
 Since $\sum_if_{\mathit{y},i}=1$ for every year $\mathit{y}=0,1,\cdots$, we have $\sum_i \kappa_{\mathit{y}+1,i}=\xi\mathcal{T}_{\mathit{y}}$. 

Now recall that annual credit amounts are distributed to households every $\Delta T$ quarters so that $\Delta T=1$ corresponds to tax credits given every quarter, $\Delta T=2$ is credits given semi-annually and $\Delta T=4$ is credits given once a year. 
 Mathematically, we can express the tax credit distributed to household $i$ in every quarter $t=4\mathit{y}+4,\cdots,4\mathit{y}+7$ of the next year $\mathit{y}+1$ as $\kappa_{t,i}=\begin{cases}
         \frac{\Delta T}{4}\cdot\kappa_{\mathit{y}+1,i}\textnormal{ if }t\%\Delta T=0\\
         0\textnormal{ otherwise }
     \end{cases}$.
 For example, assume that the government collected \$1000 as income taxes from 2 households in year $0$, and wants to distribute 10\%=\$100 as tax credits in year $1$. Also assume that it wants to give 20\%=\$20 to the first household and 80\%=\$80 to the second. If it gives credits annually with $\Delta T=4$, the households will receive \$20 and \$80 in the first quarter of the next year respectively (and zero in the remaining three quarters). On the other hand, if it gives credits quarterly with $\Delta T=1$, the households will receive \$5 and \$20 in every quarter of the next year respectively.
 
	The government optimizes its choice of tax rate and distribution of tax credits\footnote{Specifically, the government chooses fractions $f_{\mathit{y},i}$ to determine $\kappa_{t, i}$ given $\xi$ and $\Delta T$.} to maximize social welfare:
	\begin{align}\label{reward_g}
	\max_{\lbrace\tau_{t}, \{\kappa_{t, i}\}_i\rbrace_{t=0}^{H-1}} \sum_{t=0}^{H-1} \beta_{ \mathbf{G}}^t\left(\theta\cdot\mathcal{H}_t+
	\sum_{i}\frac{1}{l_{t,i}}\cdot {\kappa_{t+1,i}}\right)
	\end{align}
    where $l_{t,i}$ denotes household $i$'s liquidity at $t$ as in (\ref{eq:h_liquidity}), $\theta\in(0,1)$ is a weight for $\mathcal{H}_t$ which is the sum of household utilities at $t$, and $\beta_{\mathbf{G}}\in[0,1]$ is the government's discount factor. Social welfare is captured by the weighted sum of total household utility and inverse liquidity-weighted tax credits. Thus, the government's objective is a metric for social welfare, serving to compare and contrast different tax credit distribution schemes as illustrated in sections \ref{subsec:liquidity_g} and \ref{subsec:liquidity_lg}.

\section{Experimental Framework}
\subsection{Multi-Agent Simulator}
\begin{figure*}[t]
    \centering
    \includegraphics[width=\linewidth]{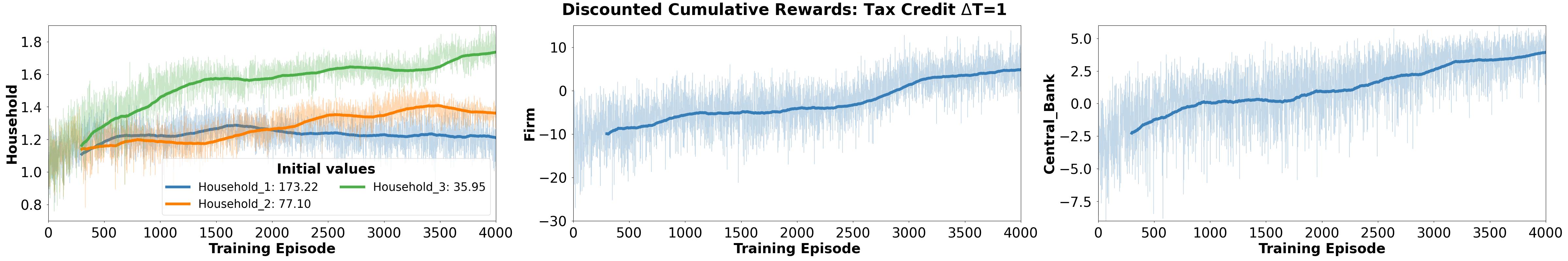}
    \includegraphics[width=\linewidth]{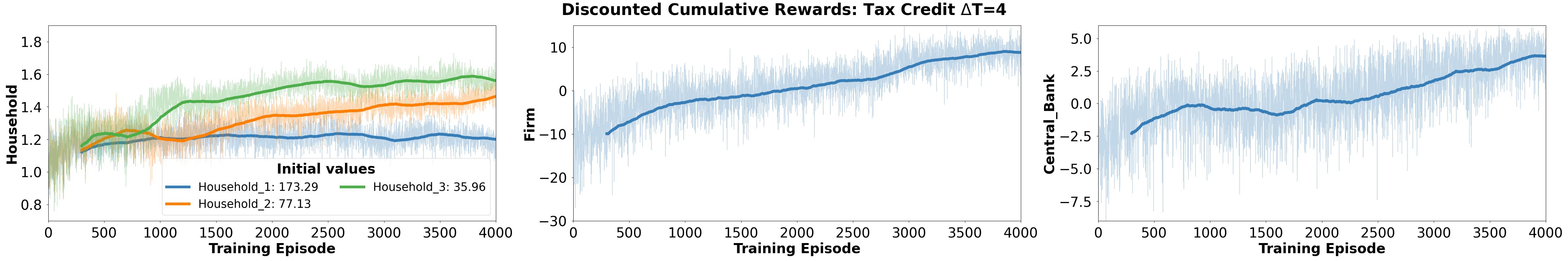}
    \caption{Training rewards with quarterly credits (first row) and annual credits (second row).}\label{fig:rewards_myopicity_g}
\end{figure*}

Our simulations are implemented using ABIDES-Economist, a multi-agent economic system simulator designed to facilitate reinforcement learning strategies for economic agents \cite{dwarakanath2024abides}. 
In ABIDES-Economist, agents can access their internal states and receive information about other agents via messages. 
A simulation kernel handles this message passing and runs simulations over a specified time horizon, maintaining timestamps for all agents and the overall simulation.
We adapt the simulator to model interactions between the economic agents described in section \ref{sec:econ_model}. Each agent is initialized with starting states and parameters informed by existing literature, particularly studies that examine real monetary policy \cite{evans2005policy}. These parameters are designed to reflect realistic economic behaviors, providing a robust foundation for agent decision-making. Agents act based on their observations, as outlined in Table \ref{table}, with their interactions driving the evolution of the economic system. 

Although our agent-based simulator is not calibrated using real economic time series data, agent action spaces are constructed to include variations around typical values observed in real data from US Bureau of Labor Statistics \shortcite{bls} e.g., labor hours chosen from $\lbrace0,240,480,720,960\rbrace$ where $480$ hours per quarter is the default action analogous to 40 hours per week. And, wages in \$ per hour are chosen from $\lbrace7.25,19.65,32.06,44.46,56.87\rbrace$ where $7.25$ is the minimum wage and $32.06$ is the default. Similarly, income tax rate choices for the government are chosen to reflect the marginal tax rates associated with the various tax brackets in the US \cite{irs}. Validation checks reported in \cite{dwarakanath2024abides} demonstrate that the simulator successfully reproduces key economic stylized facts, such as the inverse relationship between firm prices and consumption, as well as the direct relationship between inflation and interest rates. 
We further validate our simulator by qualitatively reproducing trends observed in real data used in the JPMorgan Chase report \cite{JPMC} in section \ref{subsec:liquidity_g}.

\subsection{Learning details}\label{subsec:learning_details}
	We describe the learning setup including agent actions, starting states and default parameter values unless specified otherwise.
	\begin{itemize}
		\item Households: 
  Consumption choices range from $0$ to $24$ units of goods, in increments of $6$ units. Labor choices range from $0$ to $960$ hours per quarter, in increments of $240$ hours. Parameters are $\gamma_i=0.33$, $\nu_i=0.5$ and $\mu_i=5$, with starting value $m_{0,i}=0$. 
		
		\item Firm: Price choices range from $\$188$ to $\$456$ per unit, in increments of $\$67$. Wage choices range from $\$7.25$ to $\$56.87$ per hour, in increments of $\$12.405$. Parameters are $\alpha=0.4$, $\rho=0.97$, $\sigma=0.1$, and $\chi=0.1$, with starting values $\epsilon_0=1$ and $Y_0=0$.
		
		\item Central Bank: Interest rate choices range from $0.25\%$ to $5.75\%$, in increments of $1.375\%$. Parameters are $\pi^*=1.02$ and $\lambda=0.05$. 
		
		\item Government: Income tax rate is chosen from $\lbrace10\%,12\%,22\%,24\%,$ $32\%,35\%,37\%\rbrace$. Fraction of annual credits to each household $i$, $f_{\mathit{y},i}$ is chosen from $\lbrace1,2,3,4,5\rbrace$ and then normalized by $\sum_kf_{\mathit{y},k}$. Parameters are $\xi=0.1$, $\Delta T=1$ and $\theta=0.1$. 
	\end{itemize}

Unless otherwise mentioned, all agents have discount factors $\beta_{i,\mathbf{H}}=\beta_{\mathbf{F}}=\beta_{\mathbf{CB}}=\beta_{\mathbf{CB}}=0.99$. The horizon is 10 years ($H=40$ quarters). To ease learning, we normalize agent rewards using median values for agent actions within their action spaces as follows:
	\begin{itemize}
		\item Household objective:  $\sum_{t=0}^{H-1} \beta_{i, \mathbf{H}}^t  u (c_{t, i}, \frac{n_{t, i}}{\tilde{n}}, \frac{m_{t+1, i}}{{\tilde{n}}\cdot\sum_{j}{\tilde{w}}}; \gamma_{i}, \nu_i, \mu_i )$
		where $\tilde{n}=480$ hours per quarter and $\tilde{w}=\$32.06$ per hour.
		\item Firm objective: $\sum_{t=0}^{H-1} \beta_{\mathbf{F}}^t\left(\frac{p_{t}\sum_ic_{t,i}}{\tilde{p}\sum_k\tilde{c}}-\frac{w_{t}\sum_in_{t,i}}{\tilde{w}\sum_k\tilde{n}}
    -\chi \frac{p_{t} Y_{t+1}}{\sum_i\exp(10\sigma)\tilde{p}\tilde{n}}\right)$
		where $\tilde{p}=\$322$ per unit and $\tilde{c}=12$ units.
		\item Central bank objective: $\sum_{t=0}^{H-1} \beta_{\bf{CB}}^t\left(-\left(\pi_t-\pi^{\star}\right)^2+\lambda\left( \frac{y_{t}}{ \tilde{y}}\right)^2\right)$
		where $\tilde{y} = \left(\sum_i \tilde{n}\right)^{\alpha}$.
		\item Government objective: $\sum_{t=0}^{H-1} \beta_{ \mathbf{G}}^t\left(\theta\cdot\mathcal{H}_t^{\mathrm{norm}}+\sum_{i}\frac{1}{l_{t,i}}\cdot \frac{\kappa_{t+1,i}}{\sum_k{\tilde{p}}\cdot{\tilde{c}}}\right)$
		where $\mathcal{H}_t^{\mathrm{norm}}$ is the sum of normalized household rewards at $t$.
	\end{itemize}
	The Proximal Policy Optimization algorithm in RLlib is used to independently learn agent policies \cite{rllib}, with learning rates set by grid search over $\lbrace10^{-3},2\times10^{-3},5\times10^{-3},10^{-2}\rbrace$ for each scenario.

\section{Experimental Results}
\begin{figure*}
    \centering
    \includegraphics[width=\linewidth]{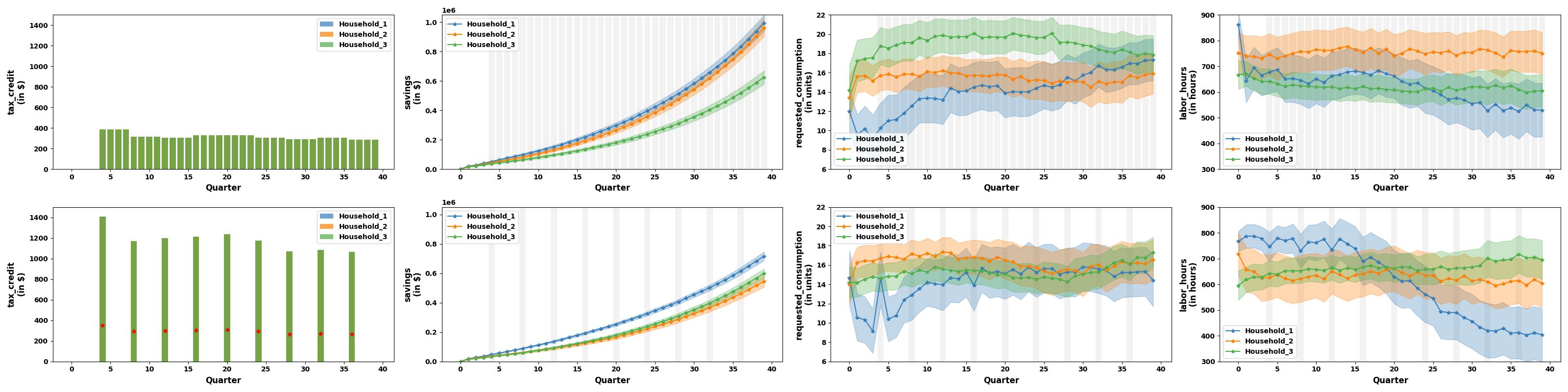}
    \caption{Household observables over time with quarterly credits (first row) and annual credits (second row).
    }
    \label{fig:myopicity_g_time}
\end{figure*}
\subsection{Impact of tax credit frequency on myopic households}\label{subsec:myopicity_g}
We investigate how households with varied degrees of myopia respond to tax credits distributed at different frequencies. We train agent policies in each credit frequency regime and test their performance in the same regime. We consider 3 households, firm and central bank as learning agents. A rule-based government agent distributes credits equally in two regimes: quarterly credits ($\Delta T=1$) and annual credits ($\Delta T=4$). Recall that discount factor $\beta_{i,\mathbf{H}}$ controls the degree of myopia of household $i$. So, household discount factors are set as $\beta_{1,\mathbf{H}}=0.99$, $\beta_{2,\mathbf{H}}=0.95$ and $\beta_{3,\mathbf{H}}=0.9$ in increasing order of myopia. 
Learning rates are $2\times10^{-3}$ for households, $5\times10^{-3}$ for the firm, and $5\times10^{-3}$ for the central bank. Discounted cumulative rewards during training are shown in Figure \ref{fig:rewards_myopicity_g}, where household rewards are plotted by normalizing to initial values due to large variations in scale across households. Shaded lines show episode rewards with solid lines showing their moving average. 

\begin{figure}[t]
    \centering
    \includegraphics[width=0.49\linewidth]{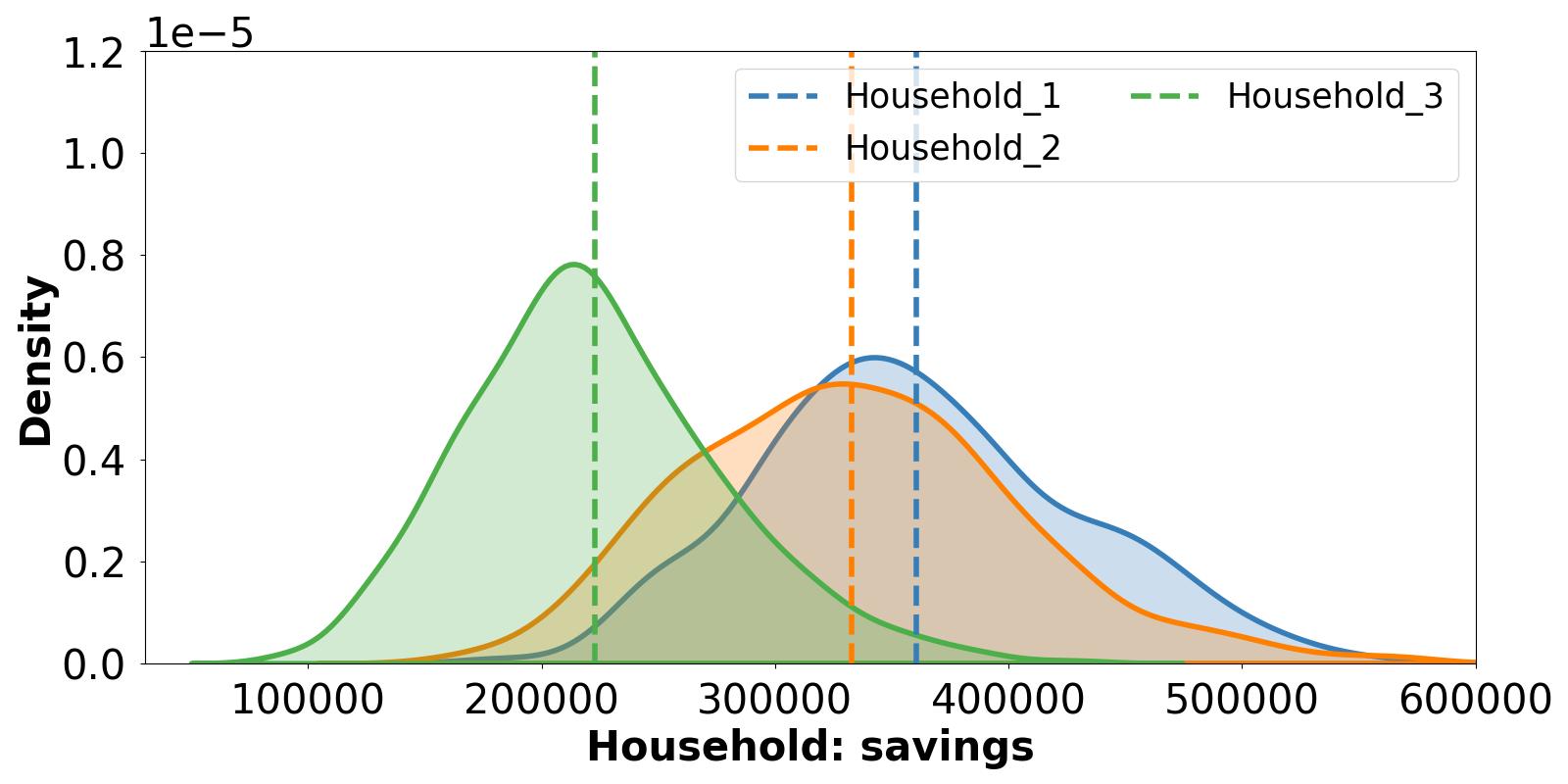}%
    \includegraphics[width=0.49\linewidth]{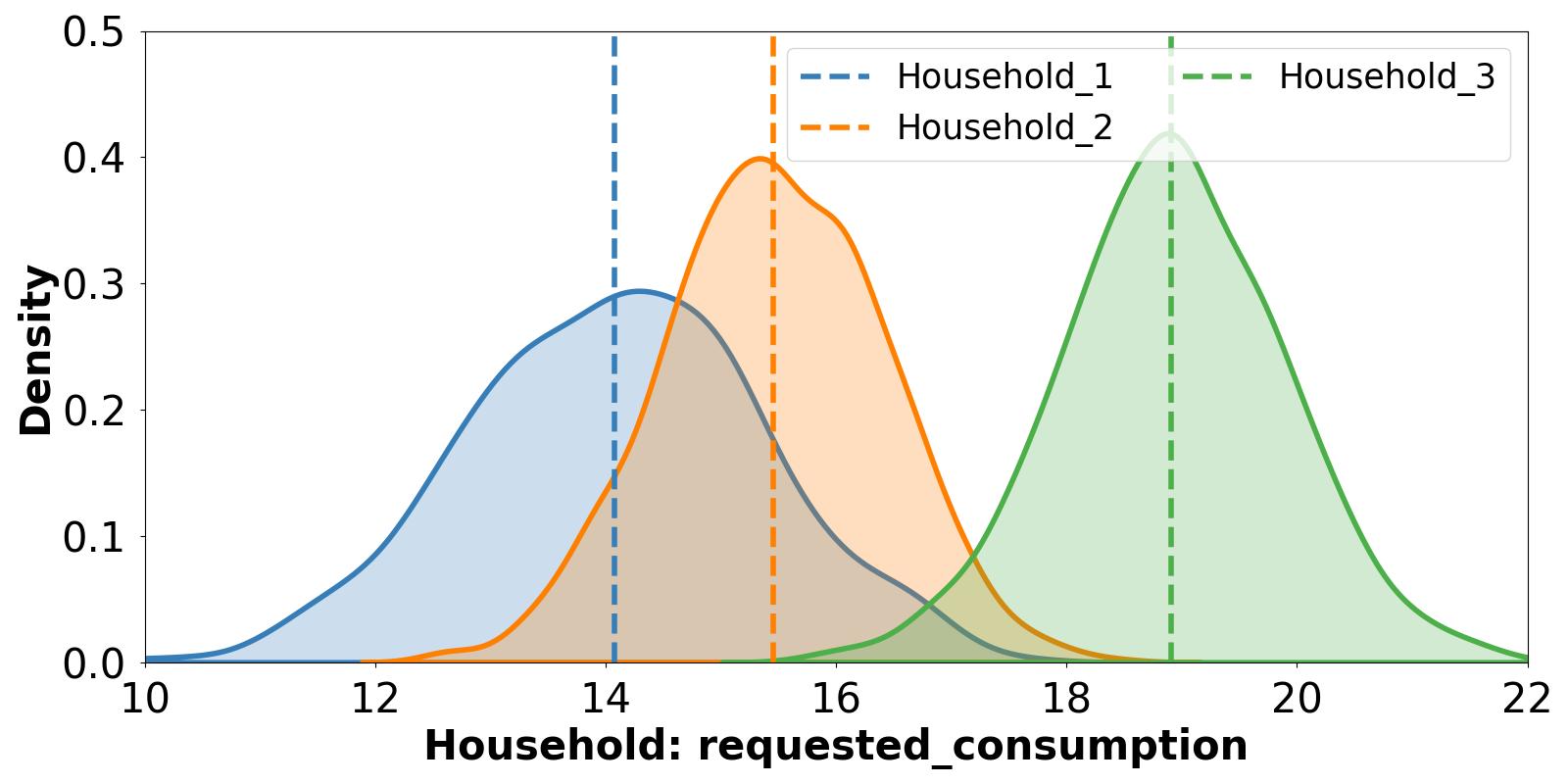}
    \includegraphics[width=0.49\linewidth]{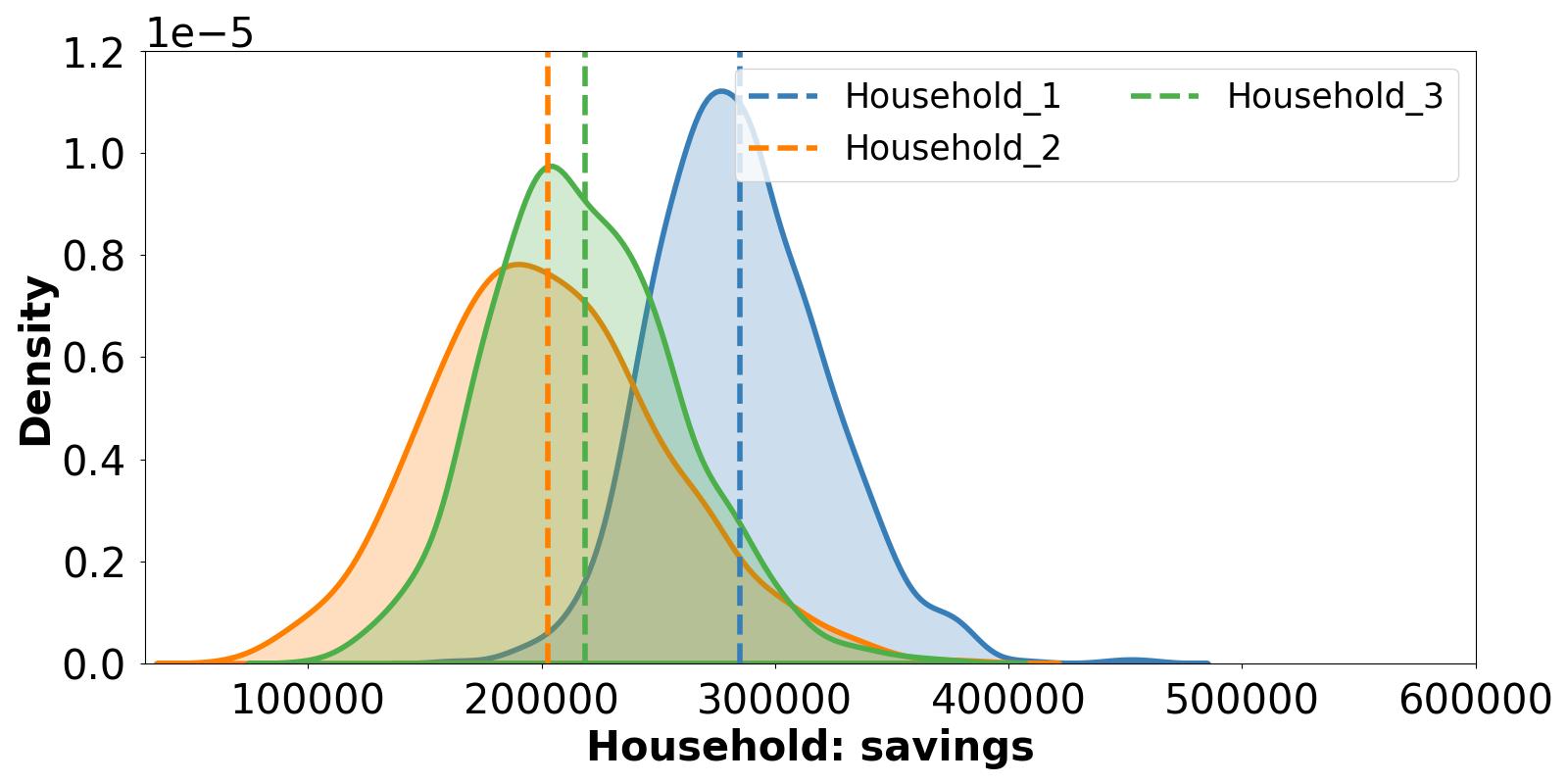}%
    \includegraphics[width=0.49\linewidth]{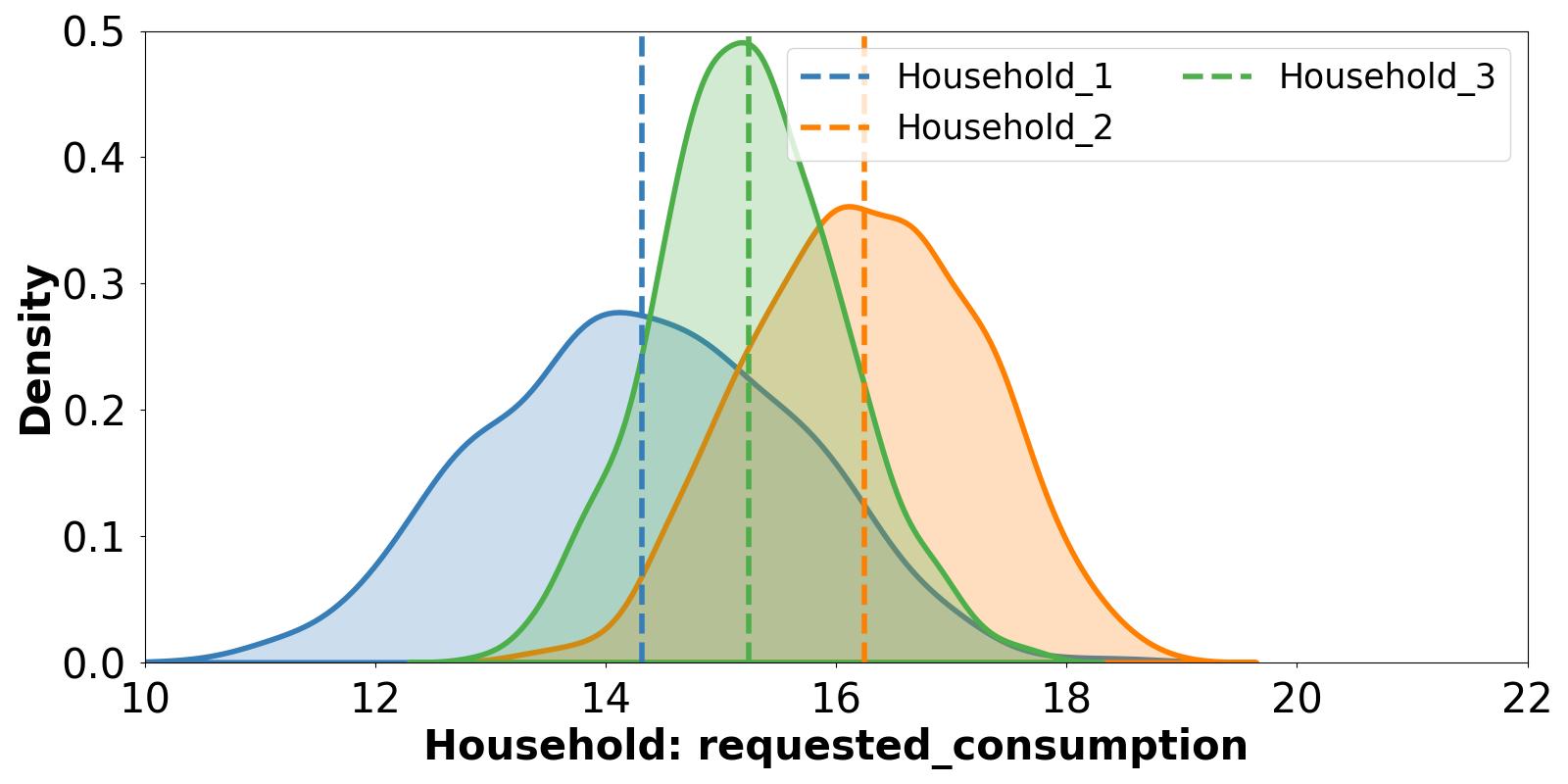}
    \caption{Distribution of average household observables with quarterly credits (first row) and annual credits (second row).
    }
    \label{fig:myopicity_g_dist}
\end{figure}

Learned policies are tested in 1000 episodes with the same training regime.
Figure \ref{fig:myopicity_g_time} shows observables versus time where solid lines represent average values across test episodes, and the shaded region is the 95\% confidence interval. The red stars in the tax credits subplot with $\Delta T=4$ are equivalent quarterly values i.e., annual value divided by four. Figure \ref{fig:myopicity_g_dist} shows the distribution across test episodes of average observations per episode. 
In both figures, the first row of plots corresponds to quarterly tax credits, where we make the following observations.
\begin{itemize}
    \item Household savings decrease with an increase in myopia as seen both from final savings at the end of the horizon from the temporal plot, and average savings from the distribution plot. This is explained by examining consumption and labor patterns below.
    \item Average household consumption increases with an increase in myopia as seen from the distribution plot. 
    Moreover, the least myopic household steadily increases its consumption over time, while more myopic households maintain relatively stable consumption at higher levels. This is because consumption utility is less discounted for the least myopic household, allowing it to consume lower and steadily increase that value as discounting increases over time. Whereas, the more myopic households need to consume higher amounts to compensate for higher discounting.
    \item Similarly, the least myopic household steadily reduces its labor hours over time as it continues receiving regular credits, while more myopic households maintain stable labor hours. The latter do not reduce labor hours continuously as labor disutility diminishes over time due to their high discounting, unlike the more sustained disutility experienced by the least myopic household.
    \item Lower initial consumption and higher initial labor of the least myopic household enable it to accumulate higher savings (amplified by interests on the same) than the most myopic household.
\end{itemize}
The second row of plots in both figures corresponds to annual tax credits, where we make the following observations.\begin{itemize}
    \item Household savings are highest for the least myopic household, while those of more myopic households are similar to one another (and smaller than former).
    \item Post the first annual credit at $t=4$, the least myopic household shows a consumption spike, in line with the notion that consumption follows cash inflow. This spike is absent for more myopic households due to their high discounting, which results in relatively higher and stable consumption levels regardless of credit frequency. Conversely, the least myopic household, being more conservative with spending, increases its immediate consumption with lump sum credits, while still preserving savings.
    \item We see periodic drops in labor hours during the quarters when annual credits are received by the least myopic household. 
    \item The least myopic household steadily increases consumption over time starting from lower values compared to more myopic households. And, it steadily decreases labor hours over time starting from a higher value compared to more myopic households. The reasoning is the same as with quarterly credits. 
    \item Difference in savings of the same household between quarterly and annual credits is higher for less myopic households 1 and 2, which see lower savings with annual credits. This is because they perceive higher utility from large credit amounts given annually causing sharp instantaneous increases in consumption and/or decreases in labor hours. In contrast, savings of the most myopic household 3 are similar under both regimes.
\end{itemize}

\subsection{Impact of unforeseen tax credits and household liquidity}\label{subsec:liquidity_g}
\begin{figure*}[t]
    \centering
    \includegraphics[width=\linewidth]{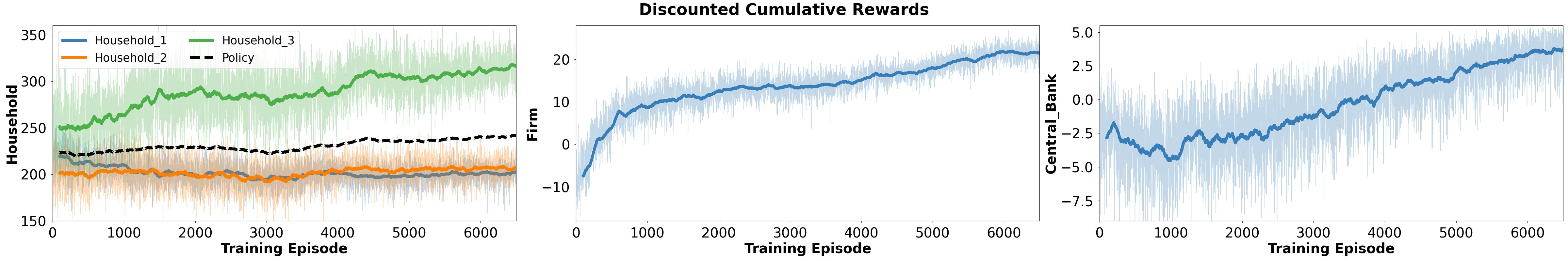}
    \caption{Training rewards without tax credits. }\label{fig:rewards_1iquidity_g}
\end{figure*}
\begin{figure*}[t]
    \centering
    \includegraphics[width=0.245\linewidth]{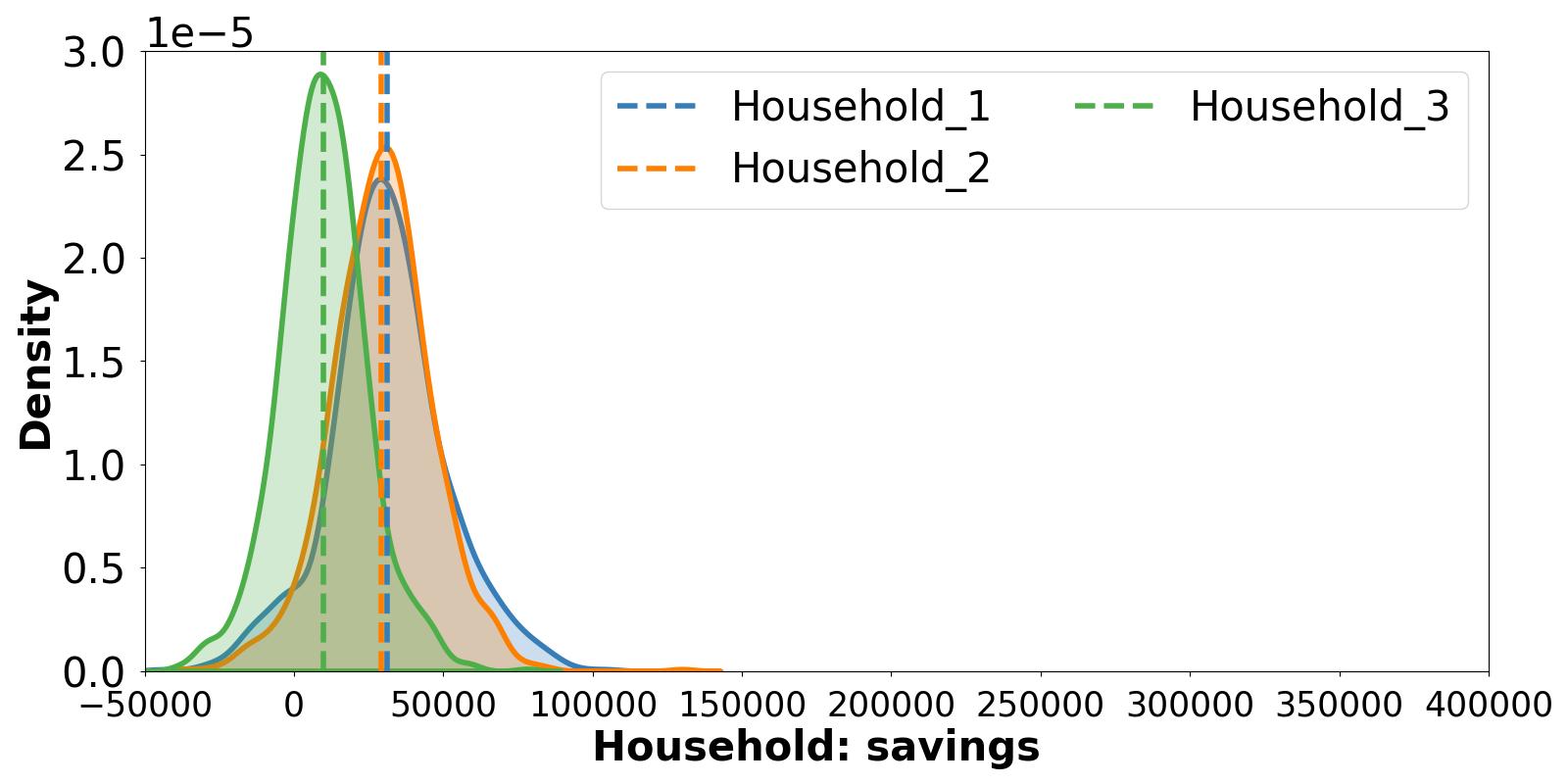}%
    \includegraphics[width=0.245\linewidth]{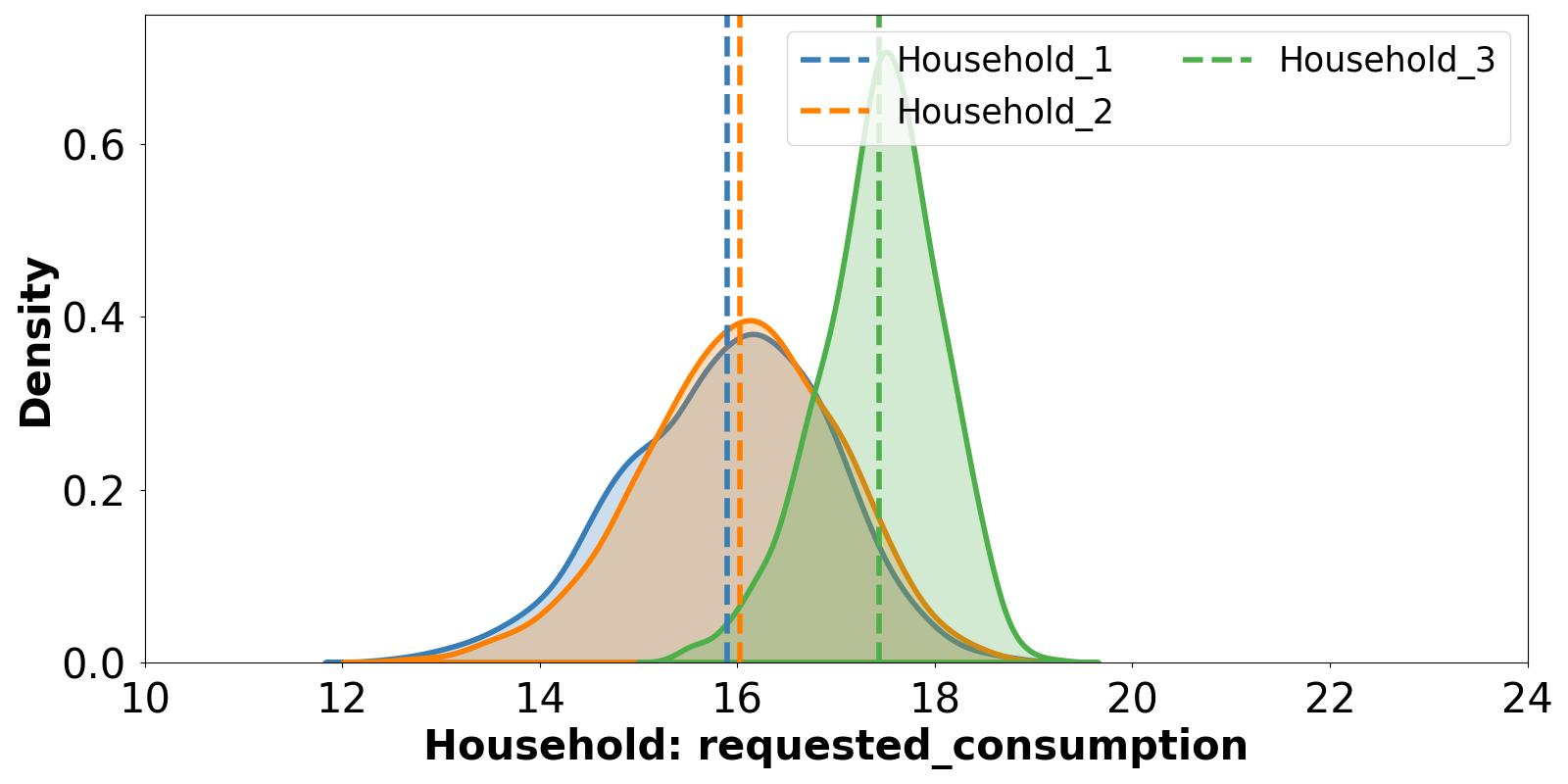}
    \includegraphics[width=0.245\linewidth]{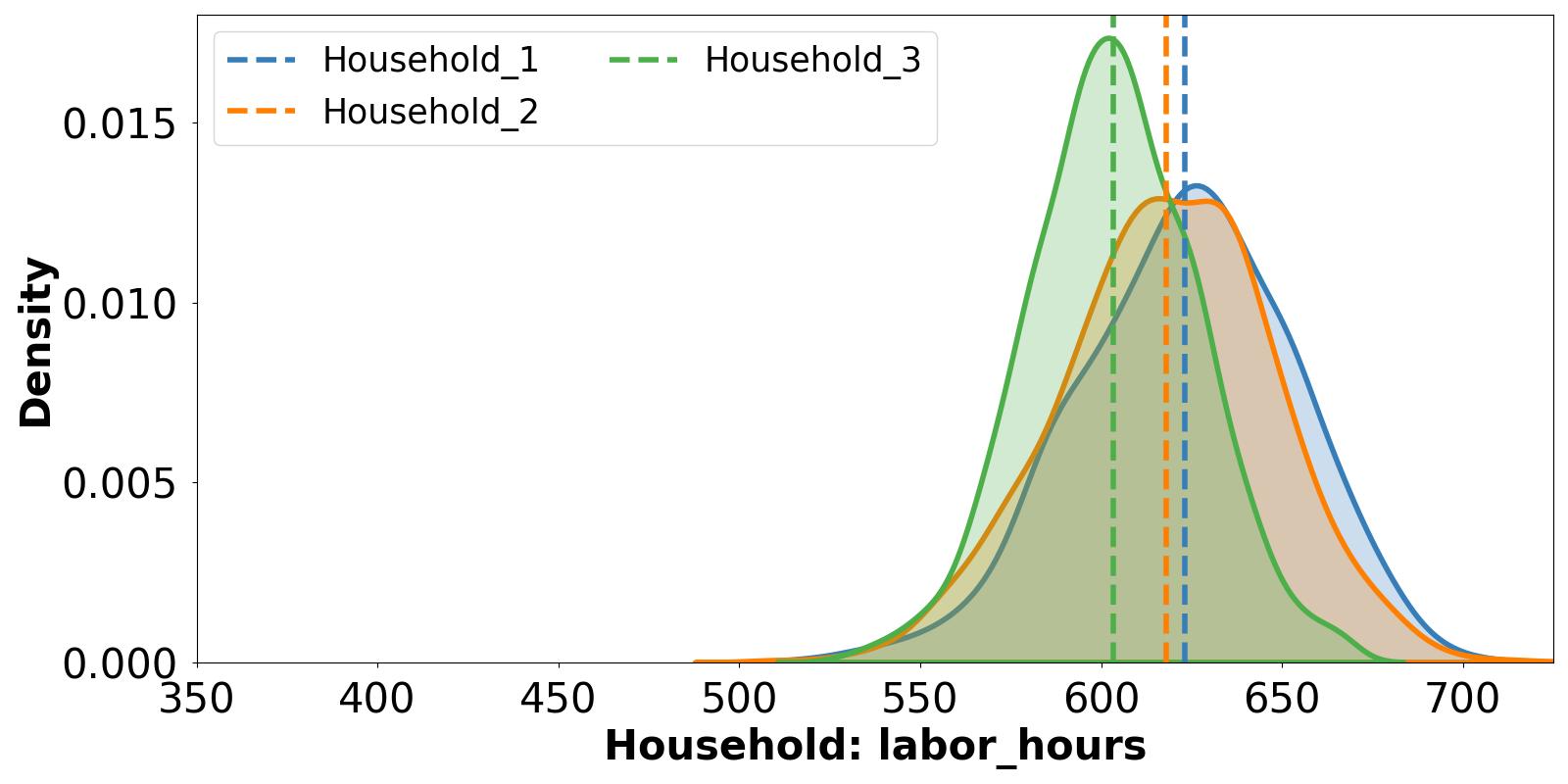}
    \includegraphics[width=0.245\linewidth]{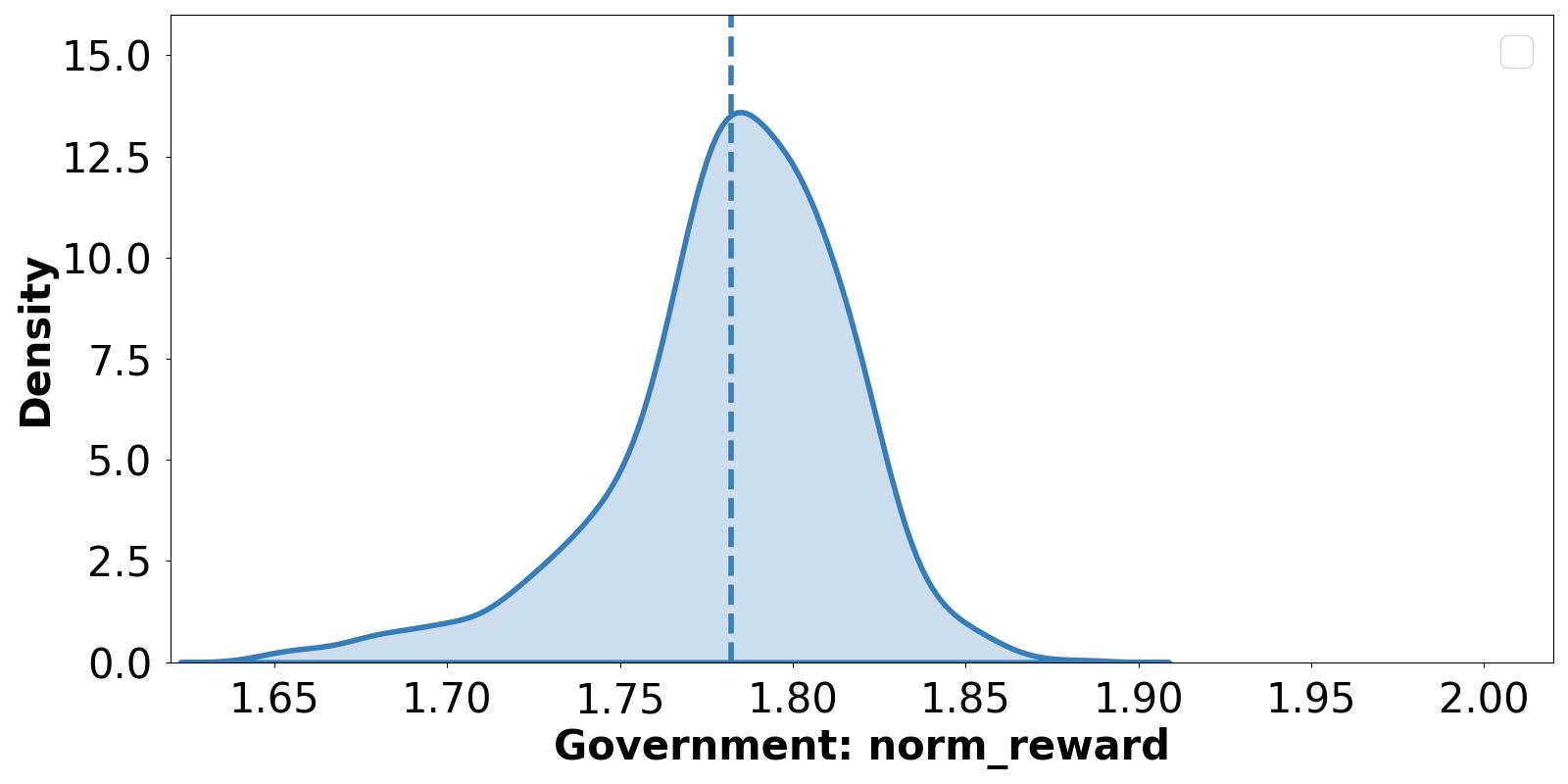}
    
    \includegraphics[width=0.245\linewidth]{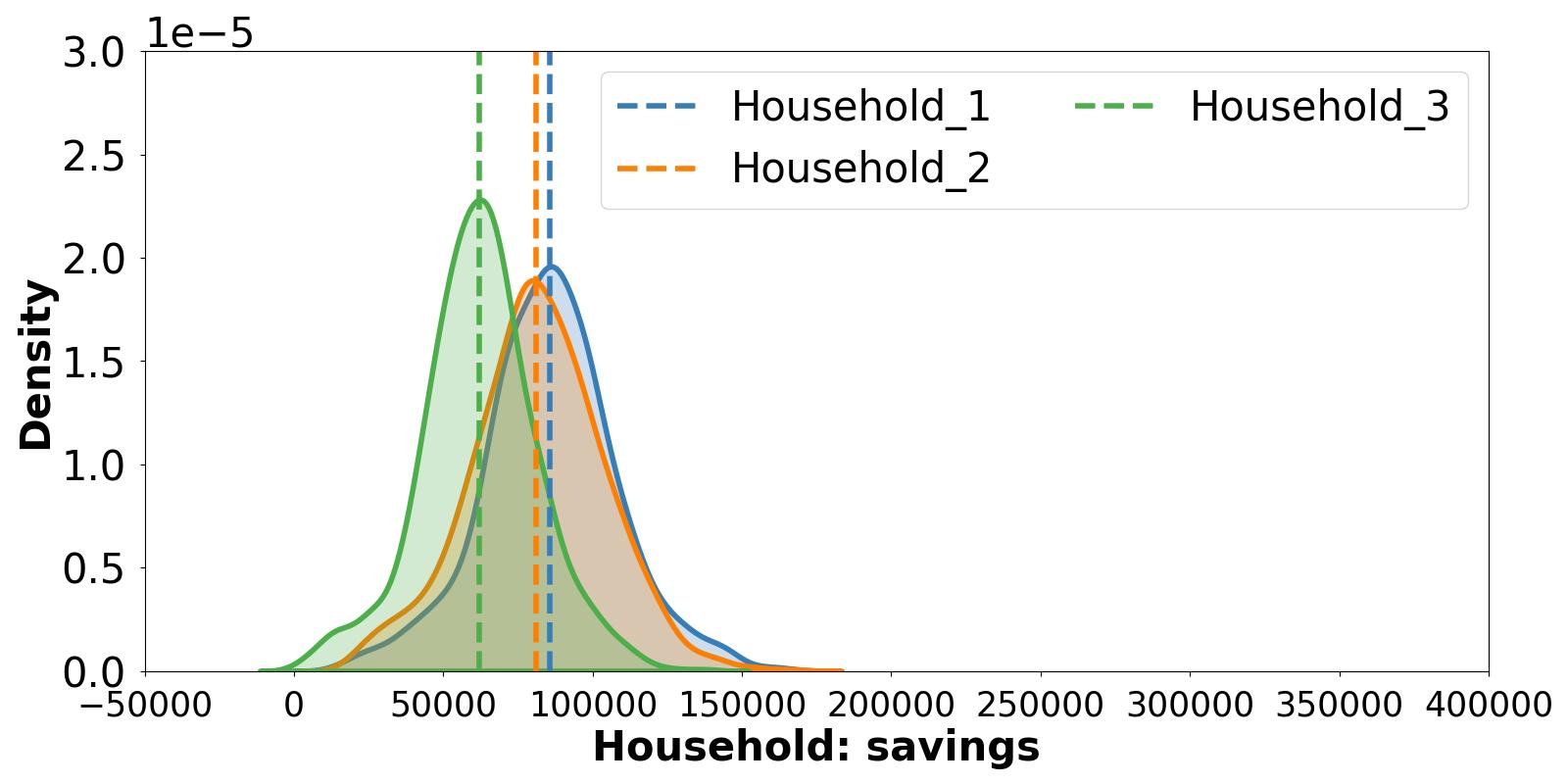}%
    \includegraphics[width=0.245\linewidth]{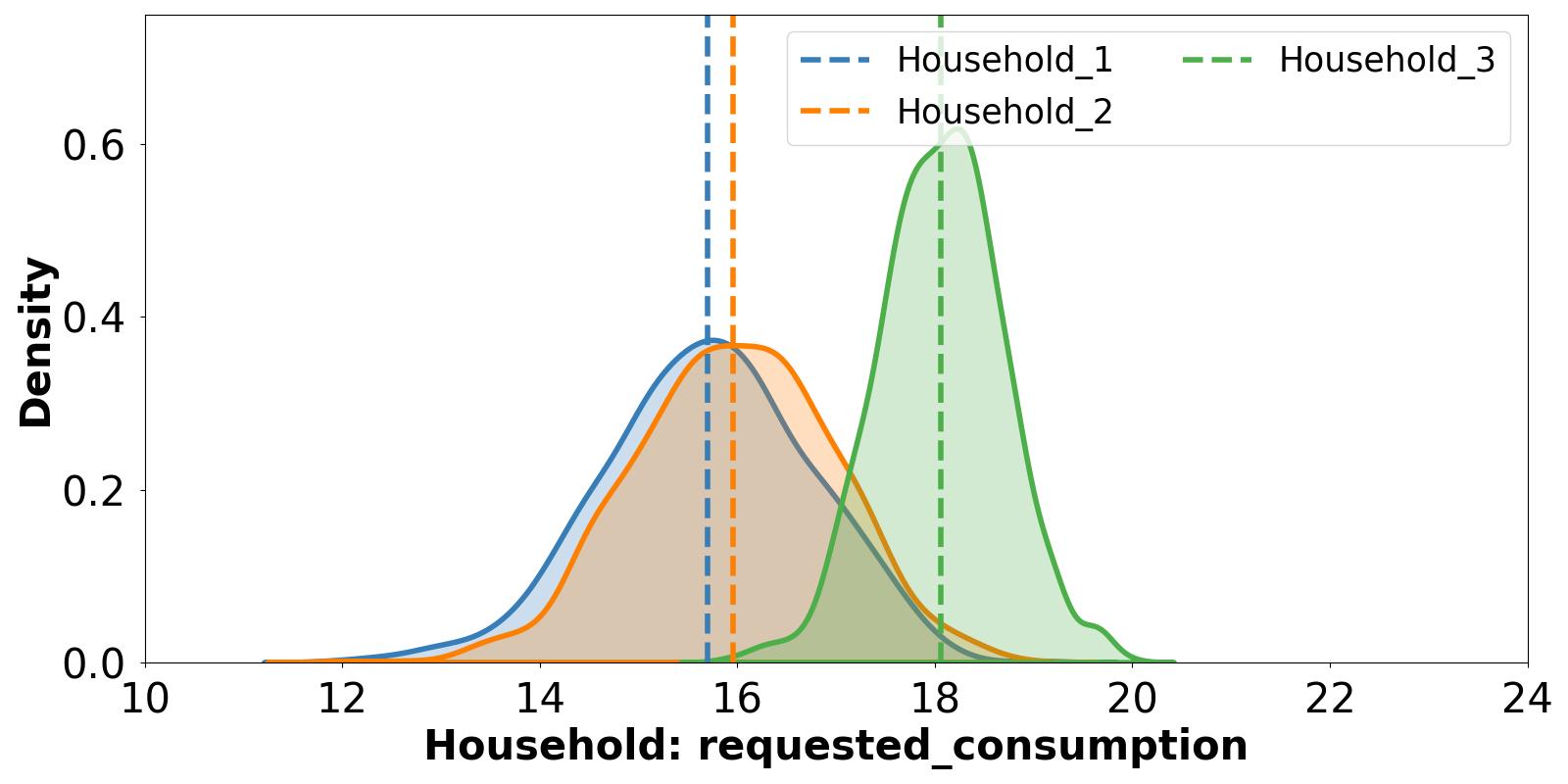}
    \includegraphics[width=0.245\linewidth]{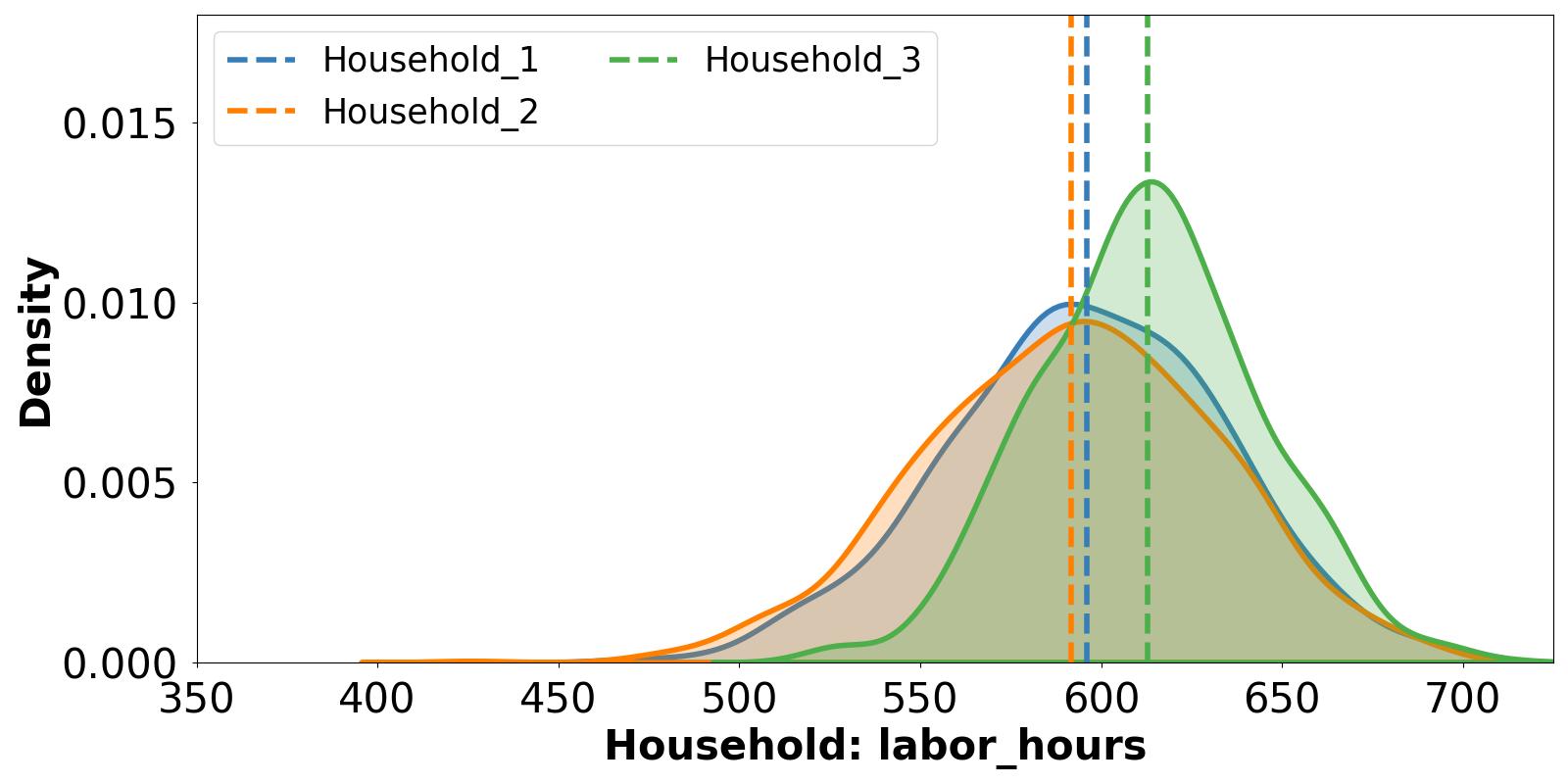}
     \includegraphics[width=0.245\linewidth]{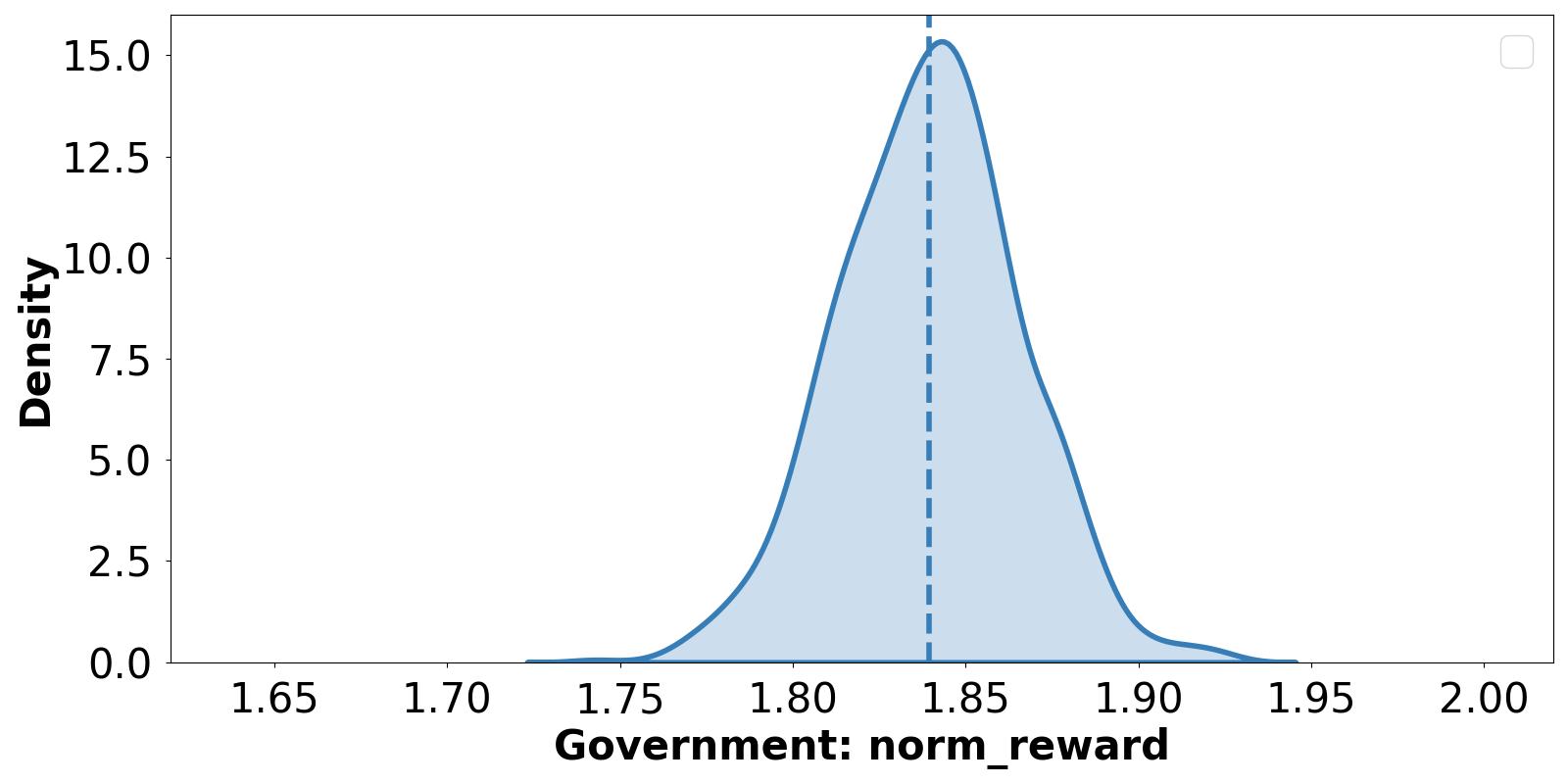}
    
    \includegraphics[width=0.245\linewidth]{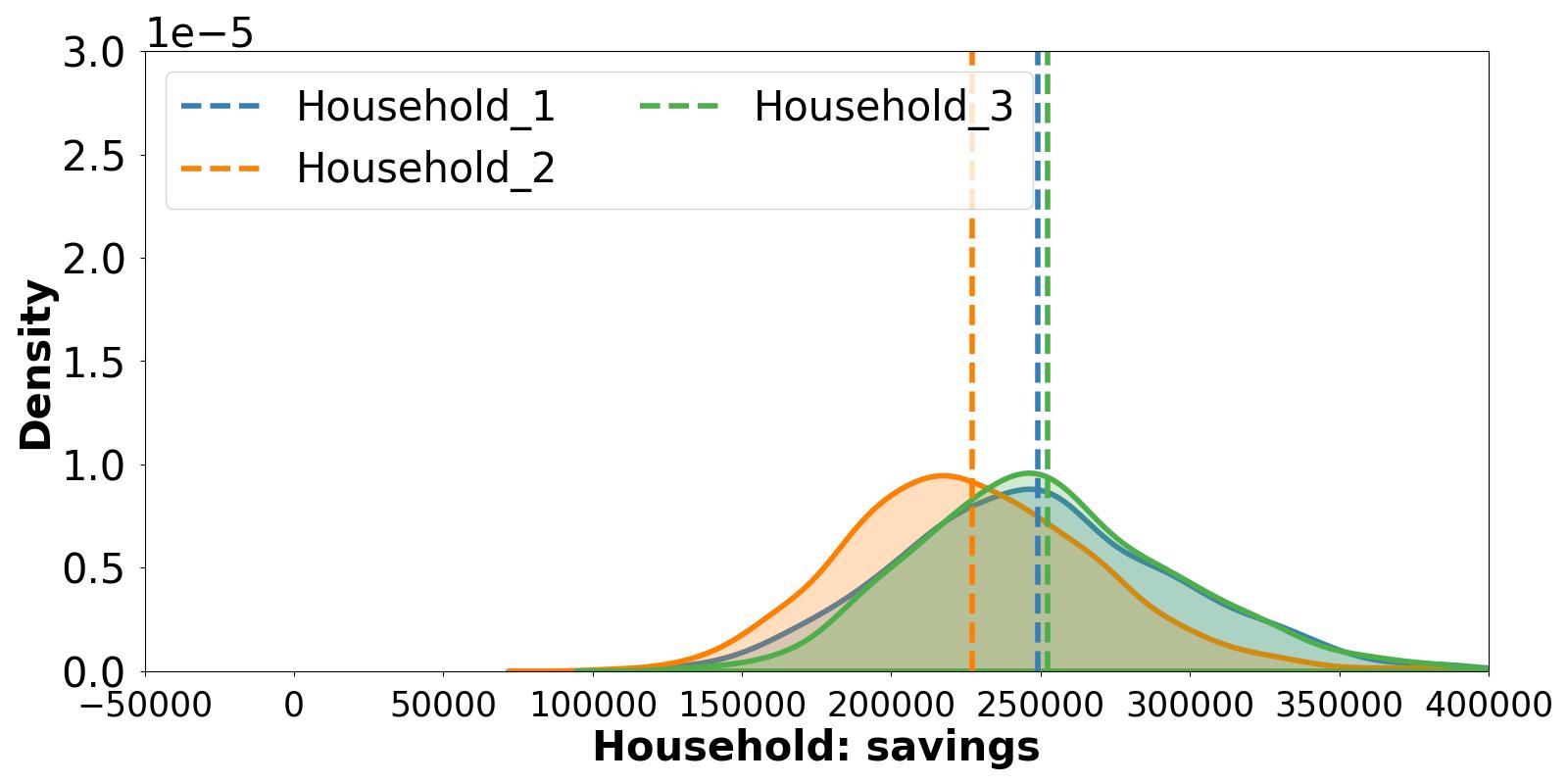}%
    \includegraphics[width=0.245\linewidth]{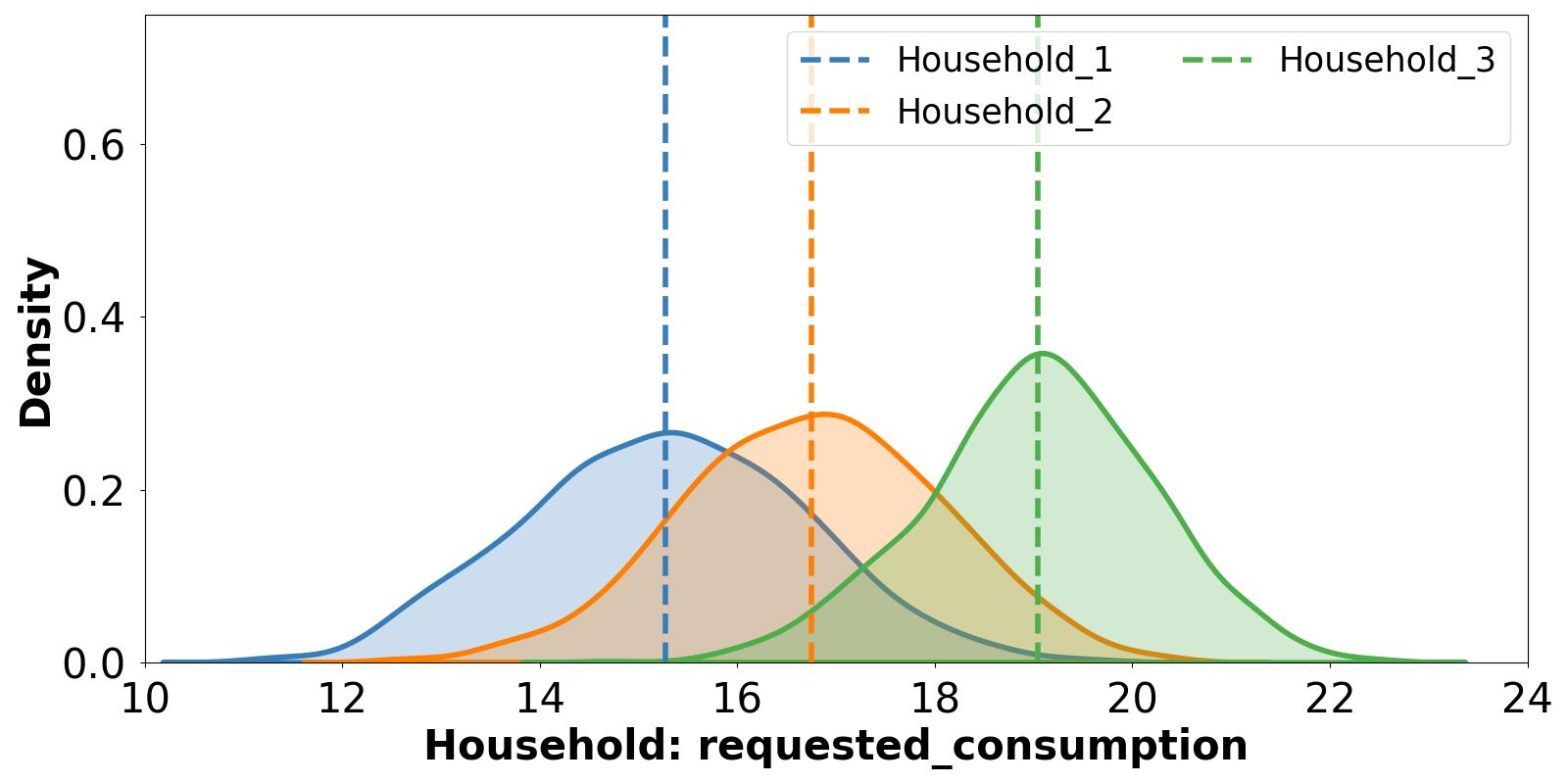}
    \includegraphics[width=0.245\linewidth]{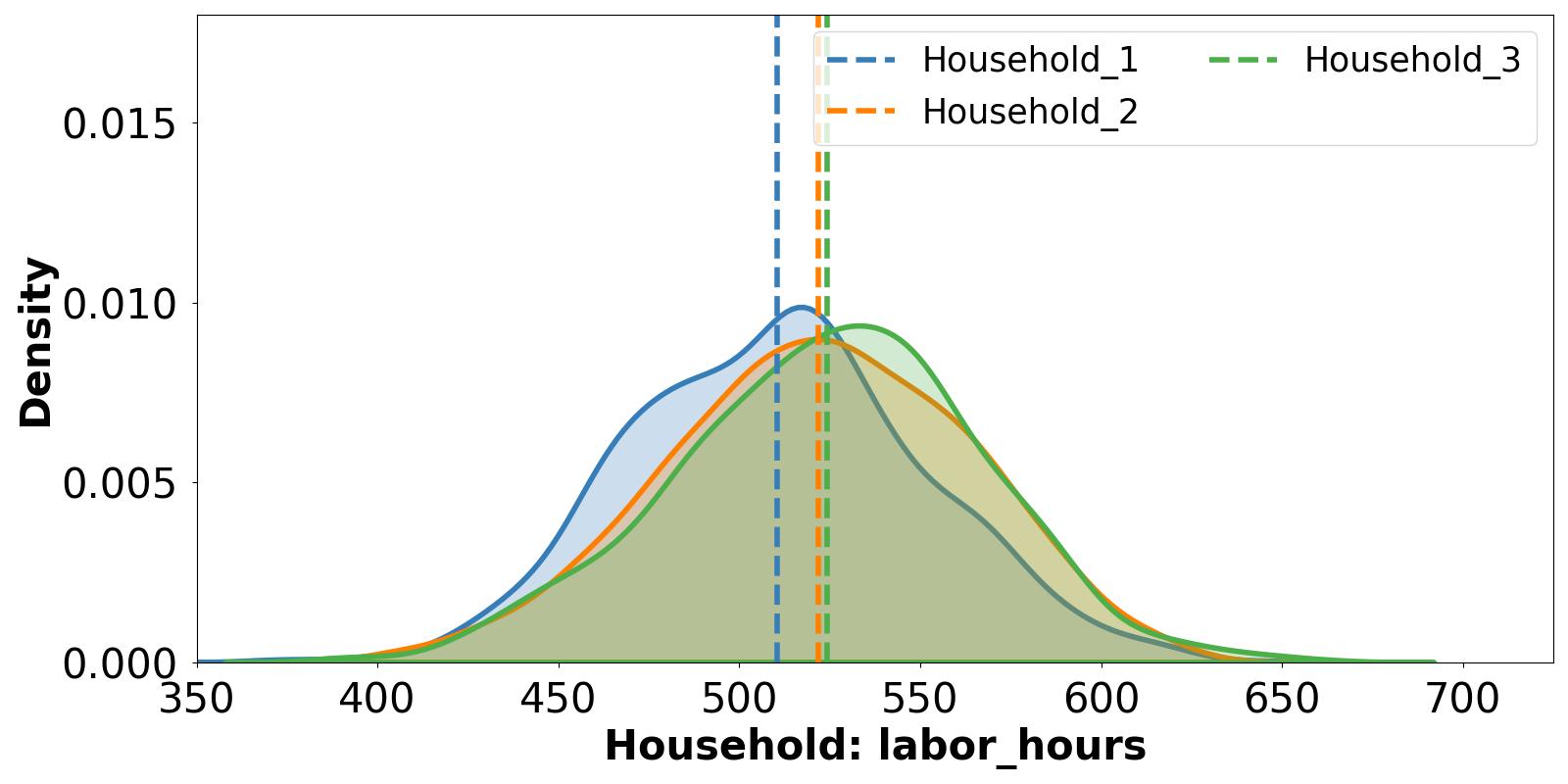}
    \includegraphics[width=0.245\linewidth]{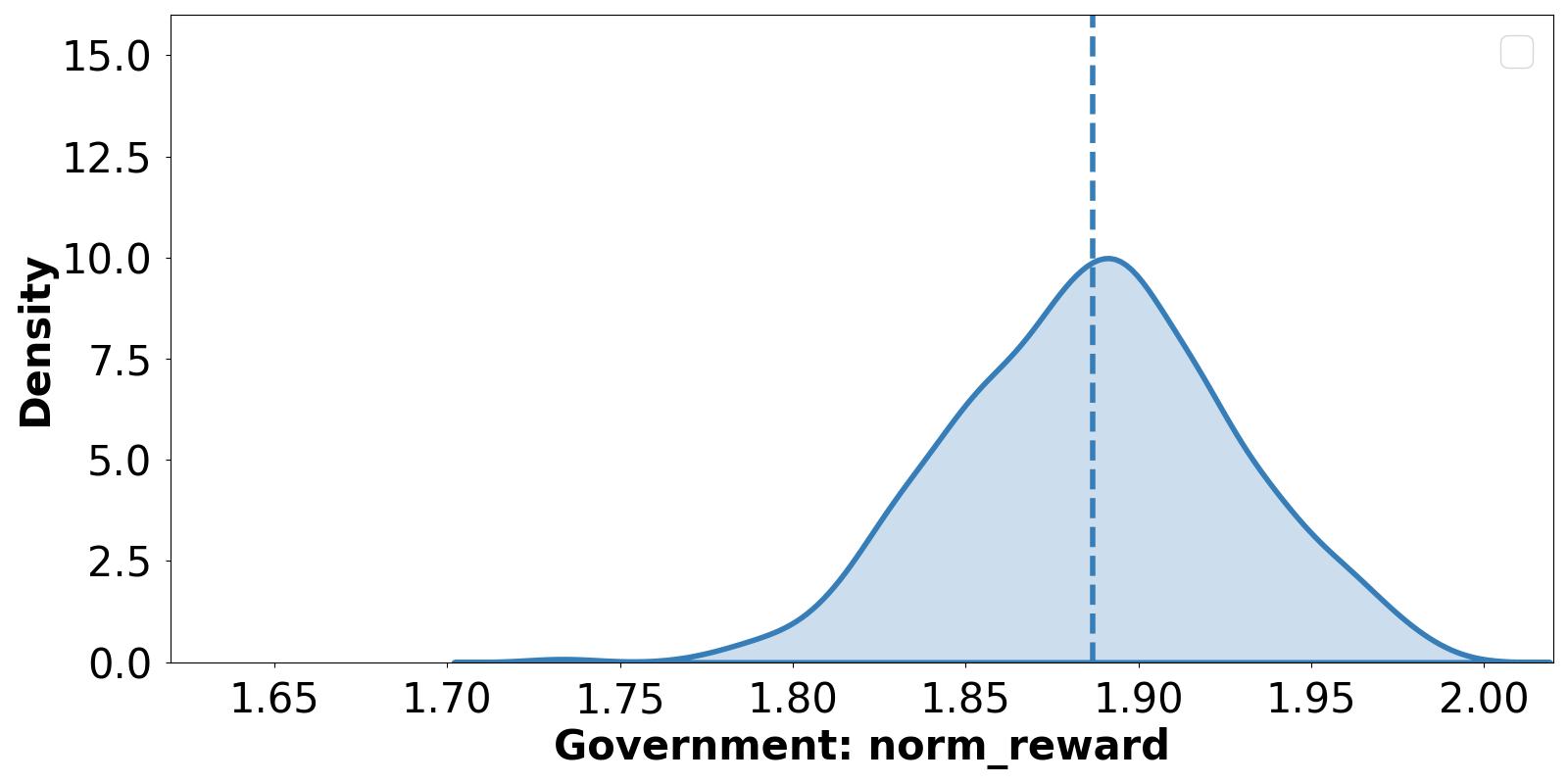}
    \caption{Household observables and government reward for social welfare in absence of tax credits (first row), presence of equal tax credits (second row) and presence of learned distribution of tax credits (third row). 
    }
    \label{fig:liquidity_g_1}
\end{figure*}

We investigate how households with different liquidities respond to unforeseen, uniform tax credits. To assess such a transitory impact of tax credits on households, we train agent policies in the absence of any credits. Learned policies are evaluated in two regimes: first, in absence of credits, and second, when the government distributes equal credits to households. Here, we consider 3 households, firm and central bank as learning agents. Also, a rule-based government agent is included to either distribute no credits ($\xi=0$) or distribute all tax revenue as credits ($\xi=1$) equally among households.

Recall that household liquidity measures savings relative to consumption spending, indicating the household's propensity to save versus consume. High liquidity households can be seen as prioritizing savings compared to low liquidity households that prioritize consumption. To reflect this tendency in our household agents, we set their utility parameters as $\gamma_1 = 0.5, \mu_1 = 5.00, \gamma_2 = 0.3, \mu_2 = 3.33,\gamma_3 = 0.1, \mu_3 = 1.67 $ in order of decreasing liquidity from Household 1 to Household 3\footnote{Notice that isoelastic consumption utility decreases as $\gamma$ goes from $\gamma=0.1$ to $\gamma=0.3$, and then to $\gamma=0.5$ for our range of consumption values specified in section \ref{subsec:learning_details}.}. We use a shared household policy network that inputs their utility parameters along with observations in Table \ref{table}.
Learning rates are $2\times10^{-3}$ for households, $5\times10^{-3}$ for the firm, and $5\times10^{-3}$ for the central bank. Figure \ref{fig:rewards_1iquidity_g} shows discounted cumulative rewards during training, where the black dashed line represents the shared household policy. 

First, learned policies are tested in 1000 episodes with no tax credits, same as during training. The first row of Figure \ref{fig:liquidity_g_1} shows the distribution of average household observables across these test episodes. Recall that household social welfare is measured by the government reward function (\ref{reward_g}). Although the government is not a learning agent in this experiment, we can measure social welfare by its reward as plotted in the last column of Figure \ref{fig:liquidity_g_1}. In absence of tax credits, household savings decrease as we go from household 1 to household 3, in decreasing order of household liquidity. And, consumption increases from household 1 to household 3. This validates our choice of heterogeneity parameters for simulating households with liquidity ranging from high to low. 

Second, learned policies are tested in 1000 episodes where the government redistributes all collected taxes as equal credits to households. The results are shown in the second row of Figure \ref{fig:liquidity_g_1}, where the same ordering of households persists in savings and consumption i.e., savings decrease and consumption increases as household liquidity decreases. With uniform credits, we see that household savings are higher than with no credits. Also, Household 3 with the lowest liquidity sees the highest increase in consumption with uniform credits, with higher liquidity households seeing little difference. This observation is in line with findings in the JPMorgan Chase report \cite{JPMC}. We also observe that labor hours reduce for higher liquidity households 1 and 2 with uniform credits while those of the lowest liquidity household 3 remain similar to the case with no credits. Importantly, household social welfare is higher with uniform tax credits than without any credits.

\subsection{Proposal for tax credit distribution scheme} \label{subsec:liquidity_lg}
\begin{figure}[t]
    \centering
    \includegraphics[width=\linewidth]{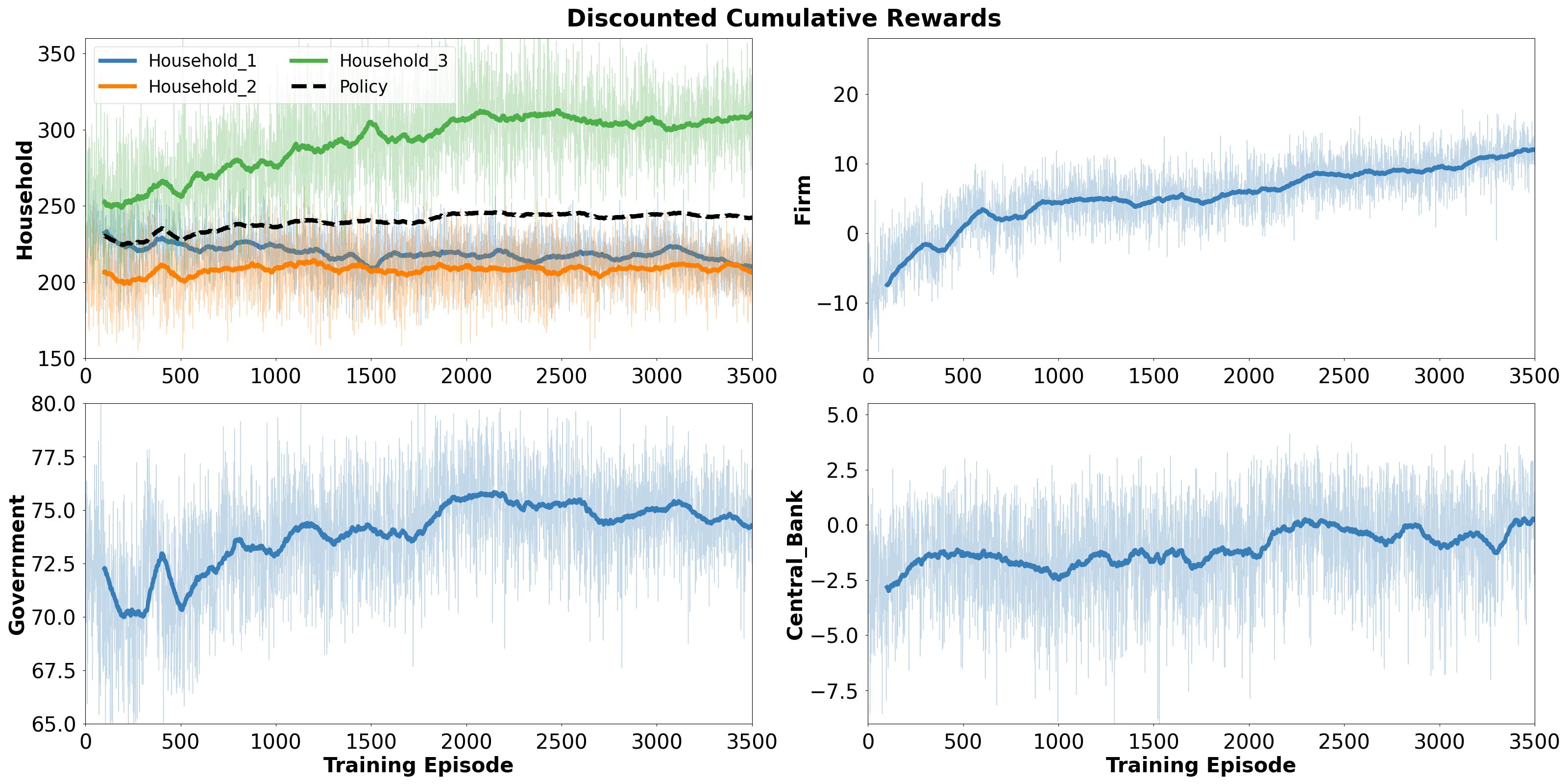}
    \caption{Training rewards with tax credits and learning government.}\label{fig:rewards_1iquidity_lg}
\end{figure}
Previously, we saw that while household savings increased with uniform tax credits, low liquidity households still had lower savings than higher liquidity ones. 
This prompted us to learn a government strategy to distribute tax credits to maximize household social welfare. We consider the same set of agents as in the above experiment with the government also being a learning agent.
It learns to distribute all tax revenue as credits ($\xi=1$) in a way to maximize its reward (i.e., social welfare) as given in (\ref{reward_g}). Learning rates are $2\times10^{-3}$ for households, $5\times10^{-3}$ for the firm, $5\times10^{-3}$ for the central bank and $10^{-2}$ for the government. Discounted cumulative rewards during training are shown in Figure \ref{fig:rewards_1iquidity_lg}. 

\begin{figure}[t]
    \centering
    \includegraphics[width=0.49\linewidth]{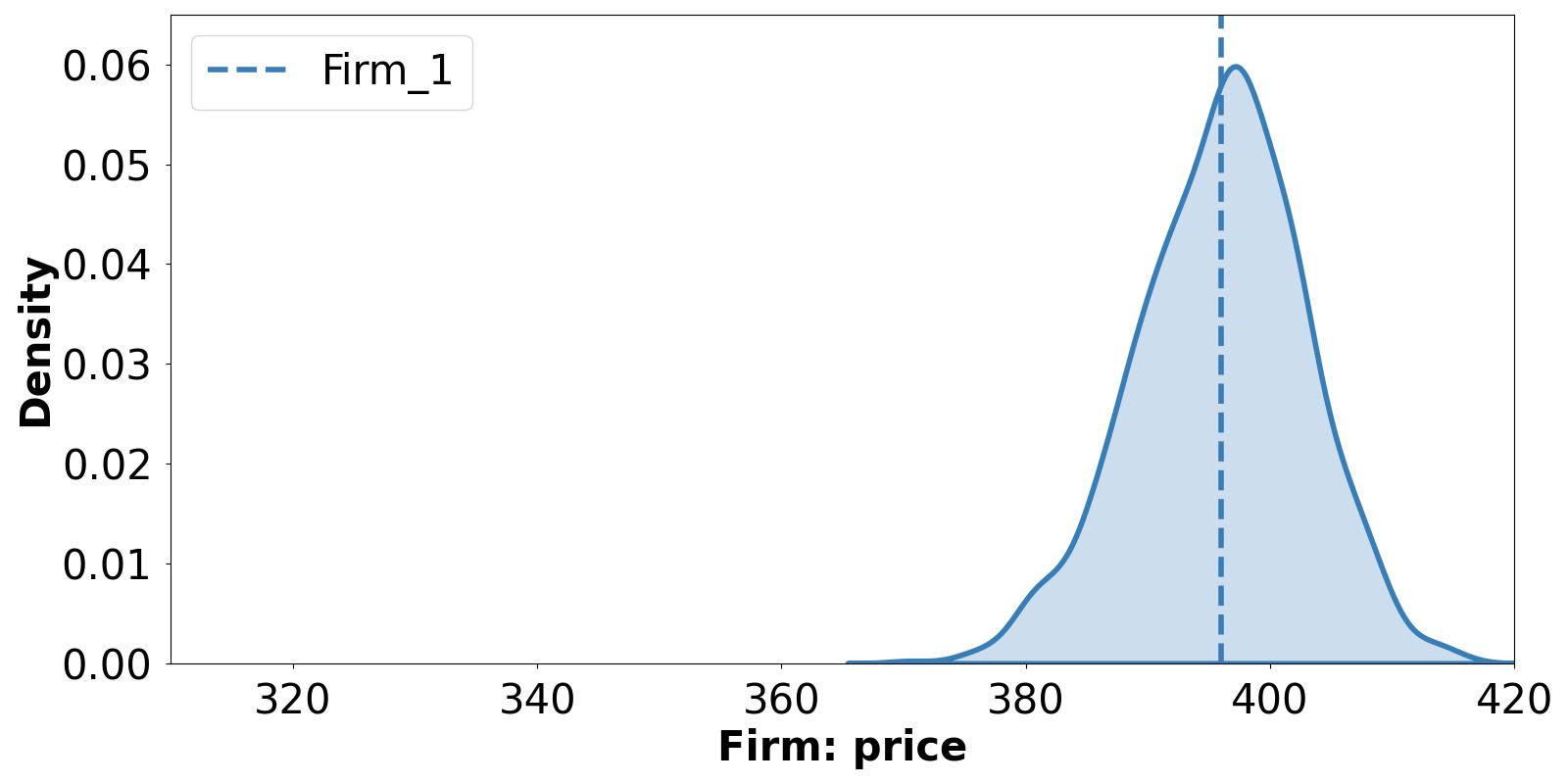}%
    \includegraphics[width=0.49\linewidth]{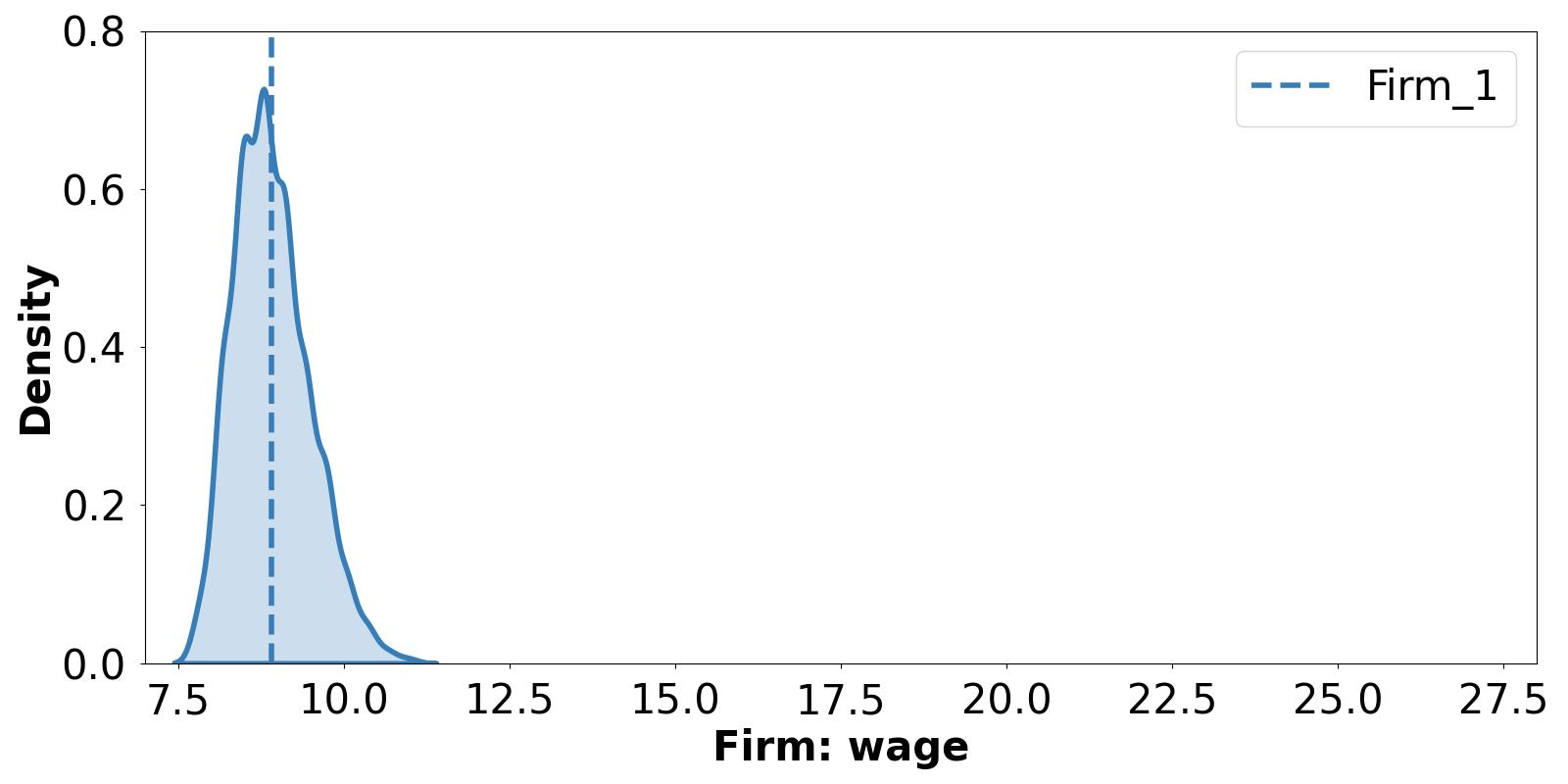}
    \includegraphics[width=0.49\linewidth]{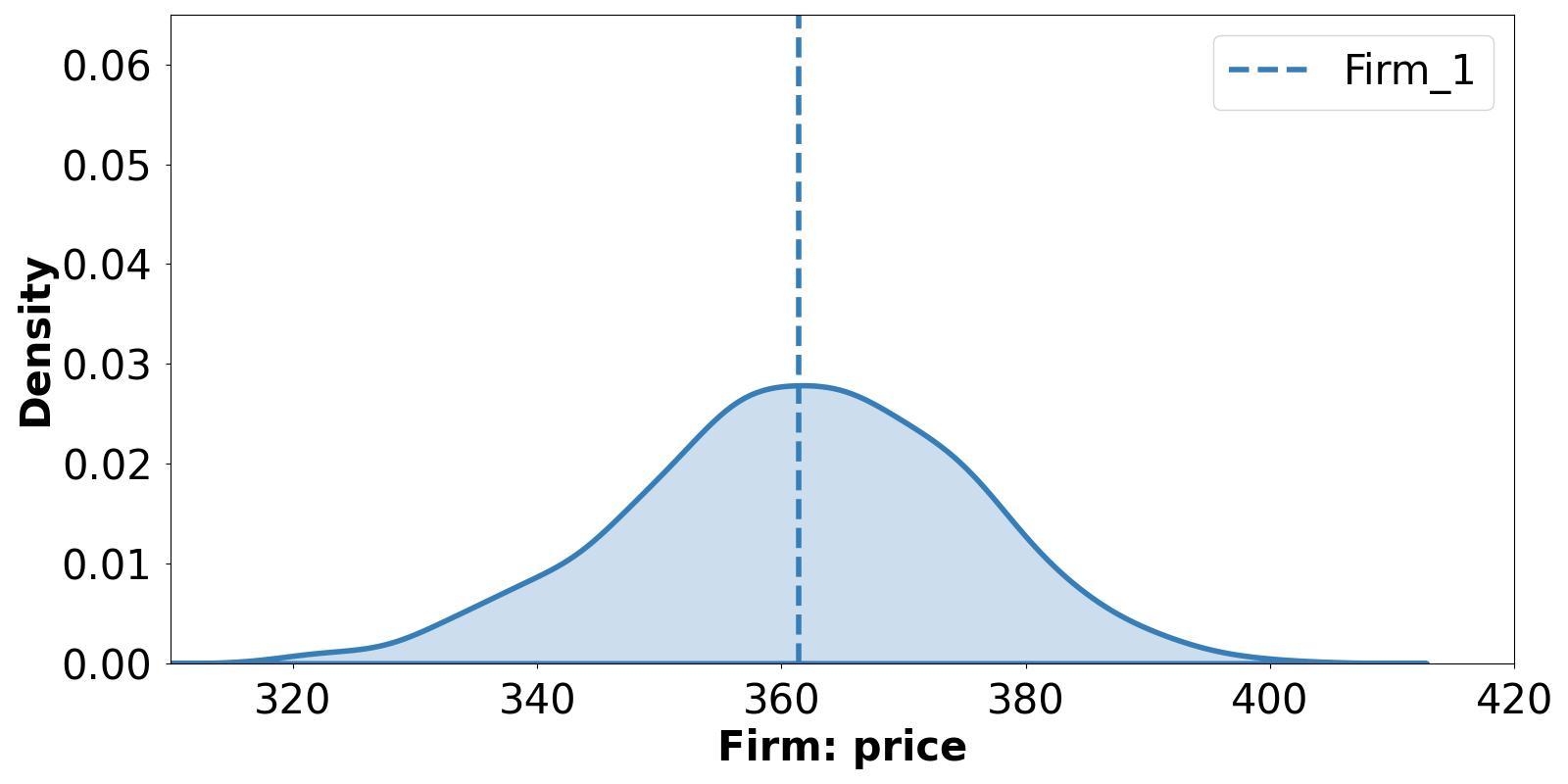}%
    \includegraphics[width=0.49\linewidth]{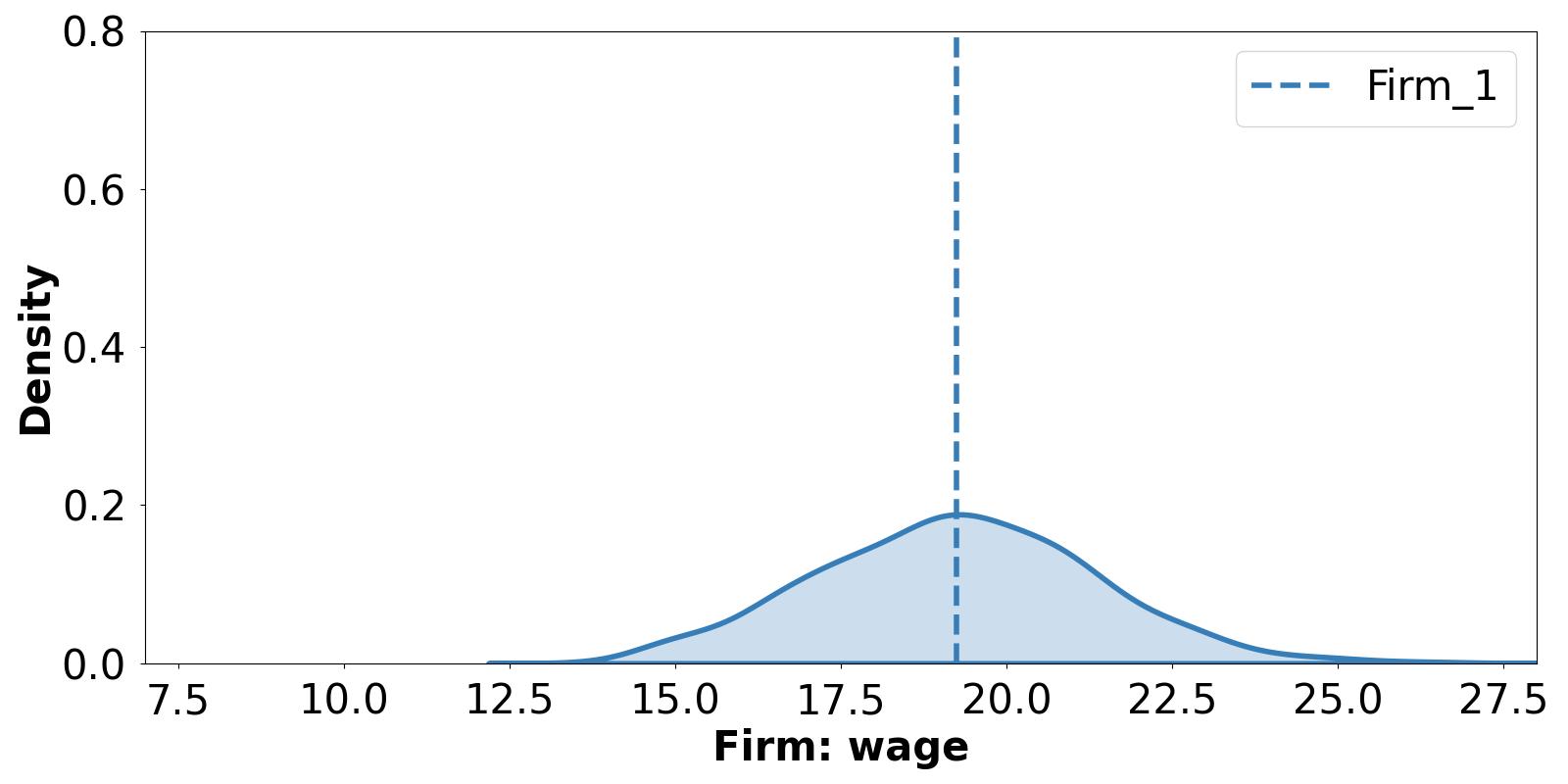}
    \caption{Distribution of average firm observables in presence of equal tax credits (first row) and presence of learned distribution of tax credits (second row).
    }
    \label{fig:liquidity_lg_firm_dist}
\end{figure}

Learned policies are tested in 1000 episodes with the same regime as during training, and results are plotted in the last row of Figure \ref{fig:liquidity_g_1}. We observe that household savings are more uniform and higher than in the previous cases, even while consumption ordering remains the same. 
To investigate this increase in savings of all households, we examine the price and wage set by the firm in Figure \ref{fig:liquidity_lg_firm_dist}. Notably, when all agents learn in presence of a government policy that adapts to maximize household social welfare, the firm learns to set lower prices and higher wages than it did under a fixed government policy.
Additionally, we see from the last column of Figure \ref{fig:liquidity_g_1} that social welfare (as measured by the government reward) is higher than before. This demonstrates the efficacy of the learned government policy that ensures equity even when households have differing propensities to consume. 

\section{Conclusion}
We investigate the impact of tax credit distribution schemes on diverse households within a simulated economy, focusing on two key aspects of household heterogeneity: myopic decision-making and liquidity. Our multi-agent economic model comprises heterogeneous households, a firm, the central bank and the government, each using reinforcement learning (RL) to optimize their strategies. 

Using the RL discount factor to model myopia, our first set of experiments examines the impact of tax credit frequency regime (annual versus quarterly) on households with varied myopia. Less myopic households display higher sensitivity to the regime, showing sharp consumption spikes and labor dips with annual credits. This results in different savings outcomes across regimes. In contrast, more myopic households exhibit similar savings under both regimes, demonstrating a relative robustness to credit frequency. 

Household liquidity, indicating the propensity to save versus consume, is modeled by parameters of savings and consumption utility. Our second set of experiments examines the impact of unforeseen, uniform tax credits on households with different liquidities and on social welfare. Low liquidity households see the largest increase in consumption upon tax credit receipt, reproducing in simulation the trends observed in real data in \cite{JPMC}. 
While uniform credits increase savings and social welfare compared to no credits, a savings gap persists between high- and low-liquidity households. 
We subsequently propose a credit distribution strategy to enhance social welfare and reduce household inequalities, demonstrating its efficacy in our experiments.
This work highlights the potential of agent-based simulators for economic policy analysis, suggesting future exploration of other social welfare metrics, such as Gini Index, and considering other household factors like demographics and location.


\section*{Disclaimer}
This paper was prepared for informational purposes in part by the Artificial Intelligence Research group of JPMorgan Chase \& Co. and its affiliates (``JP Morgan'') and is not a product of the Research Department of JP Morgan. JP Morgan makes no representation and warranty whatsoever and disclaims all liability, for the completeness, accuracy or reliability of the information contained herein. This document is not intended as investment research or investment advice, or a recommendation, offer or solicitation for the purchase or sale of any security, financial instrument, financial product or service, or to be used in any way for evaluating the merits of participating in any transaction, and shall not constitute a solicitation under any jurisdiction or to any person, if such solicitation under such jurisdiction or to such person would be unlawful.

\bibliographystyle{ACM-Reference-Format}
\bibliography{sample-base}


\begin{thebibliography}{47}


\ifx \showCODEN    \undefined \def \showCODEN     #1{\unskip}     \fi
\ifx \showDOI      \undefined \def \showDOI       #1{#1}\fi
\ifx \showISBNx    \undefined \def \showISBNx     #1{\unskip}     \fi
\ifx \showISBNxiii \undefined \def \showISBNxiii  #1{\unskip}     \fi
\ifx \showISSN     \undefined \def \showISSN      #1{\unskip}     \fi
\ifx \showLCCN     \undefined \def \showLCCN      #1{\unskip}     \fi
\ifx \shownote     \undefined \def \shownote      #1{#1}          \fi
\ifx \showarticletitle \undefined \def \showarticletitle #1{#1}   \fi
\ifx \showURL      \undefined \def \showURL       {\relax}        \fi
\providecommand\bibfield[2]{#2}
\providecommand\bibinfo[2]{#2}
\providecommand\natexlab[1]{#1}
\providecommand\showeprint[2][]{arXiv:#2}

\bibitem[Ainslie(1992)]%
        {ainslie1992picoeconomics}
\bibfield{author}{\bibinfo{person}{George Ainslie}.} \bibinfo{year}{1992}\natexlab{}.
\newblock \bibinfo{booktitle}{\emph{Picoeconomics: The strategic interaction of successive motivational states within the person}}.
\newblock \bibinfo{publisher}{Cambridge University Press}.
\newblock


\bibitem[Ananat et~al\mbox{.}(2022)]%
        {ananat2022effects}
\bibfield{author}{\bibinfo{person}{Elizabeth Ananat}, \bibinfo{person}{Benjamin Glasner}, \bibinfo{person}{Christal Hamilton}, {and} \bibinfo{person}{Zachary Parolin}.} \bibinfo{year}{2022}\natexlab{}.
\newblock \bibinfo{booktitle}{\emph{Effects of the expanded Child Tax Credit on employment outcomes: Evidence from real-world data from April to December 2021}}.
\newblock \bibinfo{type}{{T}echnical {R}eport}. \bibinfo{institution}{National Bureau of Economic Research}.
\newblock


\bibitem[Atashbar and Shi(2023)]%
        {atashbar2023ai}
\bibfield{author}{\bibinfo{person}{Tohid Atashbar} {and} \bibinfo{person}{Rui~Aruhan Shi}.} \bibinfo{year}{2023}\natexlab{}.
\newblock \bibinfo{booktitle}{\emph{AI and macroeconomic modeling: Deep reinforcement learning in an RBC model}}.
\newblock \bibinfo{publisher}{International Monetary Fund}.
\newblock


\bibitem[Axtell and Farmer(2022)]%
        {axtell2022agent}
\bibfield{author}{\bibinfo{person}{Robert~L Axtell} {and} \bibinfo{person}{J~Doyne Farmer}.} \bibinfo{year}{2022}\natexlab{}.
\newblock \showarticletitle{Agent-based modeling in economics and finance: Past, present, and future}.
\newblock \bibinfo{journal}{\emph{Journal of Economic Literature}} (\bibinfo{year}{2022}).
\newblock


\bibitem[Brusatin et~al\mbox{.}(2024)]%
        {brusatin2024simulating}
\bibfield{author}{\bibinfo{person}{Simone Brusatin}, \bibinfo{person}{Tommaso Padoan}, \bibinfo{person}{Andrea Coletta}, \bibinfo{person}{Domenico~Delli Gatti}, {and} \bibinfo{person}{Aldo Glielmo}.} \bibinfo{year}{2024}\natexlab{}.
\newblock \showarticletitle{Simulating the economic impact of rationality through reinforcement learning and agent-based modelling}.
\newblock \bibinfo{journal}{\emph{arXiv preprint arXiv:2405.02161}} (\bibinfo{year}{2024}).
\newblock


\bibitem[Byrd et~al\mbox{.}(2019)]%
        {byrd2019abides}
\bibfield{author}{\bibinfo{person}{David Byrd}, \bibinfo{person}{Maria Hybinette}, {and} \bibinfo{person}{Tucker~Hybinette Balch}.} \bibinfo{year}{2019}\natexlab{}.
\newblock \showarticletitle{Abides: Towards high-fidelity market simulation for ai research}.
\newblock \bibinfo{journal}{\emph{arXiv preprint arXiv:1904.12066}} (\bibinfo{year}{2019}).
\newblock


\bibitem[Chabris et~al\mbox{.}(2010)]%
        {chabris2010intertemporal}
\bibfield{author}{\bibinfo{person}{Christopher~F Chabris}, \bibinfo{person}{David~I Laibson}, {and} \bibinfo{person}{Jonathon~P Schuldt}.} \bibinfo{year}{2010}\natexlab{}.
\newblock \showarticletitle{Intertemporal choice}.
\newblock In \bibinfo{booktitle}{\emph{Behavioural and experimental economics}}. \bibinfo{publisher}{Springer}, \bibinfo{pages}{168--177}.
\newblock


\bibitem[Chen et~al\mbox{.}(2021)]%
        {chen2021deep}
\bibfield{author}{\bibinfo{person}{Mingli Chen}, \bibinfo{person}{Andreas Joseph}, \bibinfo{person}{Michael Kumhof}, \bibinfo{person}{Xinlei Pan}, \bibinfo{person}{Rui Shi}, {and} \bibinfo{person}{Xuan Zhou}.} \bibinfo{year}{2021}\natexlab{}.
\newblock \showarticletitle{Deep reinforcement learning in a monetary model}.
\newblock \bibinfo{journal}{\emph{arXiv preprint arXiv:2104.09368}} (\bibinfo{year}{2021}).
\newblock


\bibitem[Corinth et~al\mbox{.}(2021)]%
        {corinth2021anti}
\bibfield{author}{\bibinfo{person}{Kevin Corinth}, \bibinfo{person}{Bruce Meyer}, \bibinfo{person}{Matthew Stadnicki}, {and} \bibinfo{person}{Derek Wu}.} \bibinfo{year}{2021}\natexlab{}.
\newblock \showarticletitle{The anti-poverty, targeting, and labor supply effects of the proposed child tax credit expansion}.
\newblock \bibinfo{journal}{\emph{University of Chicago, Becker Friedman Institute for Economics Working Paper}} \bibinfo{number}{2021-115} (\bibinfo{year}{2021}).
\newblock


\bibitem[Cremer and Pestieau(2011)]%
        {cremer2011myopia}
\bibfield{author}{\bibinfo{person}{Helmuth Cremer} {and} \bibinfo{person}{Pierre Pestieau}.} \bibinfo{year}{2011}\natexlab{}.
\newblock \showarticletitle{Myopia, redistribution and pensions}.
\newblock \bibinfo{journal}{\emph{European Economic Review}} \bibinfo{volume}{55}, \bibinfo{number}{2} (\bibinfo{year}{2011}), \bibinfo{pages}{165--175}.
\newblock


\bibitem[Curry et~al\mbox{.}(2022)]%
        {curry2022analyzing}
\bibfield{author}{\bibinfo{person}{Michael Curry}, \bibinfo{person}{Alexander Trott}, \bibinfo{person}{Soham Phade}, \bibinfo{person}{Yu Bai}, {and} \bibinfo{person}{Stephan Zheng}.} \bibinfo{year}{2022}\natexlab{}.
\newblock \showarticletitle{Analyzing Micro-Founded General Equilibrium Models with Many Agents using Deep Reinforcement Learning}.
\newblock \bibinfo{journal}{\emph{arXiv preprint arXiv:2201.01163}} (\bibinfo{year}{2022}).
\newblock


\bibitem[Dawid and Gatti(2018)]%
        {dawid2018agent}
\bibfield{author}{\bibinfo{person}{Herbert Dawid} {and} \bibinfo{person}{Domenico~Delli Gatti}.} \bibinfo{year}{2018}\natexlab{}.
\newblock \showarticletitle{Agent-based macroeconomics}.
\newblock \bibinfo{journal}{\emph{Handbook of computational economics}}  \bibinfo{volume}{4} (\bibinfo{year}{2018}), \bibinfo{pages}{63--156}.
\newblock


\bibitem[Dawid et~al\mbox{.}(2016)]%
        {dawid2016heterogeneous}
\bibfield{author}{\bibinfo{person}{Herbert Dawid}, \bibinfo{person}{Philipp Harting}, \bibinfo{person}{Sander van~der Hoog}, {and} \bibinfo{person}{Michael Neugart}.} \bibinfo{year}{2016}\natexlab{}.
\newblock \showarticletitle{A heterogeneous agent macroeconomic model for policy evaluation: Improving transparency and reproducibility}.
\newblock  (\bibinfo{year}{2016}).
\newblock


\bibitem[Deissenberg et~al\mbox{.}(2008)]%
        {deissenberg2008eurace}
\bibfield{author}{\bibinfo{person}{Christophe Deissenberg}, \bibinfo{person}{Sander Van Der~Hoog}, {and} \bibinfo{person}{Herbert Dawid}.} \bibinfo{year}{2008}\natexlab{}.
\newblock \showarticletitle{EURACE: A massively parallel agent-based model of the European economy}.
\newblock \bibinfo{journal}{\emph{Applied mathematics and computation}} \bibinfo{volume}{204}, \bibinfo{number}{2} (\bibinfo{year}{2008}), \bibinfo{pages}{541--552}.
\newblock


\bibitem[Dosi et~al\mbox{.}(2015)]%
        {dosi2015fiscal}
\bibfield{author}{\bibinfo{person}{Giovanni Dosi}, \bibinfo{person}{Giorgio Fagiolo}, \bibinfo{person}{Mauro Napoletano}, \bibinfo{person}{Andrea Roventini}, {and} \bibinfo{person}{Tania Treibich}.} \bibinfo{year}{2015}\natexlab{}.
\newblock \showarticletitle{Fiscal and monetary policies in complex evolving economies}.
\newblock \bibinfo{journal}{\emph{Journal of Economic Dynamics and Control}}  \bibinfo{volume}{52} (\bibinfo{year}{2015}), \bibinfo{pages}{166--189}.
\newblock


\bibitem[Dosi et~al\mbox{.}(2006)]%
        {dosi2006evolutionary}
\bibfield{author}{\bibinfo{person}{Giovanni Dosi}, \bibinfo{person}{Giorgio Fagiolo}, {and} \bibinfo{person}{Andrea Roventini}.} \bibinfo{year}{2006}\natexlab{}.
\newblock \showarticletitle{An evolutionary model of endogenous business cycles}.
\newblock \bibinfo{journal}{\emph{Computational Economics}}  \bibinfo{volume}{27} (\bibinfo{year}{2006}), \bibinfo{pages}{3--34}.
\newblock


\bibitem[Dosi et~al\mbox{.}(2017)]%
        {dosi2017micro}
\bibfield{author}{\bibinfo{person}{Giovanni Dosi}, \bibinfo{person}{Mauro Napoletano}, \bibinfo{person}{Andrea Roventini}, {and} \bibinfo{person}{Tania Treibich}.} \bibinfo{year}{2017}\natexlab{}.
\newblock \showarticletitle{Micro and macro policies in the Keynes+ Schumpeter evolutionary models}.
\newblock \bibinfo{journal}{\emph{Journal of Evolutionary Economics}}  \bibinfo{volume}{27} (\bibinfo{year}{2017}), \bibinfo{pages}{63--90}.
\newblock


\bibitem[Dosi and Roventini(2019)]%
        {dosi2019more}
\bibfield{author}{\bibinfo{person}{Giovanni Dosi} {and} \bibinfo{person}{Andrea Roventini}.} \bibinfo{year}{2019}\natexlab{}.
\newblock \showarticletitle{More is different... and complex! the case for agent-based macroeconomics}.
\newblock \bibinfo{journal}{\emph{Journal of Evolutionary Economics}}  \bibinfo{volume}{29} (\bibinfo{year}{2019}), \bibinfo{pages}{1--37}.
\newblock


\bibitem[Dwarakanath et~al\mbox{.}(2024)]%
        {dwarakanath2024abides}
\bibfield{author}{\bibinfo{person}{Kshama Dwarakanath}, \bibinfo{person}{Svitlana Vyetrenko}, \bibinfo{person}{Peyman Tavallali}, {and} \bibinfo{person}{Tucker Balch}.} \bibinfo{year}{2024}\natexlab{}.
\newblock \showarticletitle{ABIDES-Economist: Agent-Based Simulation of Economic Systems with Learning Agents}.
\newblock \bibinfo{journal}{\emph{arXiv preprint arXiv:2402.09563}} (\bibinfo{year}{2024}).
\newblock


\bibitem[Ericson and Laibson(2019)]%
        {ericson2019intertemporal}
\bibfield{author}{\bibinfo{person}{Keith~Marzilli Ericson} {and} \bibinfo{person}{David Laibson}.} \bibinfo{year}{2019}\natexlab{}.
\newblock \showarticletitle{Intertemporal choice}.
\newblock In \bibinfo{booktitle}{\emph{Handbook of behavioral economics: Applications and foundations 1}}. Vol.~\bibinfo{volume}{2}. \bibinfo{publisher}{Elsevier}, \bibinfo{pages}{1--67}.
\newblock


\bibitem[Evans and Honkapohja(2005)]%
        {evans2005policy}
\bibfield{author}{\bibinfo{person}{George~W Evans} {and} \bibinfo{person}{Seppo Honkapohja}.} \bibinfo{year}{2005}\natexlab{}.
\newblock \showarticletitle{Policy interaction, expectations and the liquidity trap}.
\newblock \bibinfo{journal}{\emph{Review of Economic Dynamics}} \bibinfo{volume}{8}, \bibinfo{number}{2} (\bibinfo{year}{2005}), \bibinfo{pages}{303--323}.
\newblock


\bibitem[Farrell et~al\mbox{.}(2019)]%
        {JPMC_tax_time}
\bibfield{author}{\bibinfo{person}{Diana Farrell}, \bibinfo{person}{Fiona Greig}, {and} \bibinfo{person}{Hamoudi Amar}.} \bibinfo{year}{2019}\natexlab{}.
\newblock \bibinfo{booktitle}{\emph{Tax Time: How Families Manage Tax Refunds and Payments.}}
\newblock \bibinfo{type}{{T}echnical {R}eport}. \bibinfo{institution}{JPMorgan Chase Institute}.
\newblock


\bibitem[Feldstein(1985)]%
        {feldstein1985optimal}
\bibfield{author}{\bibinfo{person}{Martin Feldstein}.} \bibinfo{year}{1985}\natexlab{}.
\newblock \showarticletitle{The optimal level of social security benefits}.
\newblock \bibinfo{journal}{\emph{The Quarterly Journal of Economics}} \bibinfo{volume}{100}, \bibinfo{number}{2} (\bibinfo{year}{1985}), \bibinfo{pages}{303--320}.
\newblock


\bibitem[Green et~al\mbox{.}(1994)]%
        {green1994discounting}
\bibfield{author}{\bibinfo{person}{Leonard Green}, \bibinfo{person}{Astrid~F Fry}, {and} \bibinfo{person}{Joel Myerson}.} \bibinfo{year}{1994}\natexlab{}.
\newblock \showarticletitle{Discounting of delayed rewards: A life-span comparison}.
\newblock \bibinfo{journal}{\emph{Psychological science}} \bibinfo{volume}{5}, \bibinfo{number}{1} (\bibinfo{year}{1994}), \bibinfo{pages}{33--36}.
\newblock


\bibitem[Green et~al\mbox{.}(1996)]%
        {green1996temporal}
\bibfield{author}{\bibinfo{person}{Leonard Green}, \bibinfo{person}{Joel Myerson}, \bibinfo{person}{David Lichtman}, \bibinfo{person}{Suzanne Rosen}, {and} \bibinfo{person}{Astrid Fry}.} \bibinfo{year}{1996}\natexlab{}.
\newblock \showarticletitle{Temporal discounting in choice between delayed rewards: the role of age and income.}
\newblock \bibinfo{journal}{\emph{Psychology and aging}} \bibinfo{volume}{11}, \bibinfo{number}{1} (\bibinfo{year}{1996}), \bibinfo{pages}{79}.
\newblock


\bibitem[Haldane and Turrell(2019)]%
        {haldane2019drawing}
\bibfield{author}{\bibinfo{person}{Andrew~G Haldane} {and} \bibinfo{person}{Arthur~E Turrell}.} \bibinfo{year}{2019}\natexlab{}.
\newblock \showarticletitle{Drawing on different disciplines: macroeconomic agent-based models}.
\newblock \bibinfo{journal}{\emph{Journal of Evolutionary Economics}}  \bibinfo{volume}{29} (\bibinfo{year}{2019}), \bibinfo{pages}{39--66}.
\newblock


\bibitem[Hill et~al\mbox{.}(2021)]%
        {hill2021solving}
\bibfield{author}{\bibinfo{person}{Edward Hill}, \bibinfo{person}{Marco Bardoscia}, {and} \bibinfo{person}{Arthur Turrell}.} \bibinfo{year}{2021}\natexlab{}.
\newblock \showarticletitle{Solving heterogeneous general equilibrium economic models with deep reinforcement learning}.
\newblock \bibinfo{journal}{\emph{arXiv preprint arXiv:2103.16977}} (\bibinfo{year}{2021}).
\newblock


\bibitem[Hinterlang and T{\"a}nzer(2021)]%
        {hinterlang2021optimal}
\bibfield{author}{\bibinfo{person}{Natascha Hinterlang} {and} \bibinfo{person}{Alina T{\"a}nzer}.} \bibinfo{year}{2021}\natexlab{}.
\newblock \showarticletitle{Optimal monetary policy using reinforcement learning}.
\newblock \bibinfo{journal}{\emph{Deutsche Bundesbank Discussion Paper}} (\bibinfo{year}{2021}).
\newblock


\bibitem[{Internal Revenue Service}(2023)]%
        {irs}
\bibfield{author}{\bibinfo{person}{{Internal Revenue Service}}.} \bibinfo{year}{2023}\natexlab{}.
\newblock \bibinfo{title}{2023 Tax Rate Schedules}.
\newblock \bibinfo{howpublished}{\url{https://www.irs.gov/media/166986}}.
\newblock


\bibitem[Kaplan et~al\mbox{.}(2018)]%
        {kaplan2018monetary}
\bibfield{author}{\bibinfo{person}{Greg Kaplan}, \bibinfo{person}{Benjamin Moll}, {and} \bibinfo{person}{Giovanni~L Violante}.} \bibinfo{year}{2018}\natexlab{}.
\newblock \showarticletitle{Monetary policy according to HANK}.
\newblock \bibinfo{journal}{\emph{American Economic Review}} \bibinfo{volume}{108}, \bibinfo{number}{3} (\bibinfo{year}{2018}), \bibinfo{pages}{697--743}.
\newblock


\bibitem[Kaplow(2015)]%
        {kaplow2015myopia}
\bibfield{author}{\bibinfo{person}{Louis Kaplow}.} \bibinfo{year}{2015}\natexlab{}.
\newblock \showarticletitle{Myopia and the effects of social security and capital taxation on labor supply}.
\newblock \bibinfo{journal}{\emph{National Tax Journal}} \bibinfo{volume}{68}, \bibinfo{number}{1} (\bibinfo{year}{2015}), \bibinfo{pages}{7--32}.
\newblock


\bibitem[Koster et~al\mbox{.}(2022)]%
        {koster2022human}
\bibfield{author}{\bibinfo{person}{Raphael Koster}, \bibinfo{person}{Jan Balaguer}, \bibinfo{person}{Andrea Tacchetti}, \bibinfo{person}{Ari Weinstein}, \bibinfo{person}{Tina Zhu}, \bibinfo{person}{Oliver Hauser}, \bibinfo{person}{Duncan Williams}, \bibinfo{person}{Lucy Campbell-Gillingham}, \bibinfo{person}{Phoebe Thacker}, \bibinfo{person}{Matthew Botvinick}, {et~al\mbox{.}}} \bibinfo{year}{2022}\natexlab{}.
\newblock \showarticletitle{Human-centred mechanism design with Democratic AI}.
\newblock \bibinfo{journal}{\emph{Nature Human Behaviour}} \bibinfo{volume}{6}, \bibinfo{number}{10} (\bibinfo{year}{2022}), \bibinfo{pages}{1398--1407}.
\newblock


\bibitem[Krusell and Smith(1998)]%
        {krusell1998income}
\bibfield{author}{\bibinfo{person}{Per Krusell} {and} \bibinfo{person}{Anthony~A Smith, Jr}.} \bibinfo{year}{1998}\natexlab{}.
\newblock \showarticletitle{Income and wealth heterogeneity in the macroeconomy}.
\newblock \bibinfo{journal}{\emph{Journal of political Economy}} \bibinfo{volume}{106}, \bibinfo{number}{5} (\bibinfo{year}{1998}), \bibinfo{pages}{867--896}.
\newblock


\bibitem[Laibson(1997)]%
        {laibson1997golden}
\bibfield{author}{\bibinfo{person}{David Laibson}.} \bibinfo{year}{1997}\natexlab{}.
\newblock \showarticletitle{Golden eggs and hyperbolic discounting}.
\newblock \bibinfo{journal}{\emph{The Quarterly Journal of Economics}} \bibinfo{volume}{112}, \bibinfo{number}{2} (\bibinfo{year}{1997}), \bibinfo{pages}{443--478}.
\newblock


\bibitem[Liang et~al\mbox{.}(2018)]%
        {rllib}
\bibfield{author}{\bibinfo{person}{Eric Liang}, \bibinfo{person}{Richard Liaw}, \bibinfo{person}{Robert Nishihara}, \bibinfo{person}{Philipp Moritz}, \bibinfo{person}{Roy Fox}, \bibinfo{person}{Ken Goldberg}, \bibinfo{person}{Joseph Gonzalez}, \bibinfo{person}{Michael Jordan}, {and} \bibinfo{person}{Ion Stoica}.} \bibinfo{year}{2018}\natexlab{}.
\newblock \showarticletitle{RLlib: Abstractions for Distributed Reinforcement Learning}. In \bibinfo{booktitle}{\emph{International Conference on Machine Learning}}.
\newblock


\bibitem[Liu et~al\mbox{.}(2022a)]%
        {liu2022biased}
\bibfield{author}{\bibinfo{person}{Penghang Liu}, \bibinfo{person}{Kshama Dwarakanath}, {and} \bibinfo{person}{Svitlana~S Vyetrenko}.} \bibinfo{year}{2022}\natexlab{a}.
\newblock \showarticletitle{Biased or limited: Modeling sub-rational human investors in financial markets}.
\newblock \bibinfo{journal}{\emph{arXiv preprint arXiv:2210.08569}} (\bibinfo{year}{2022}).
\newblock


\bibitem[Liu et~al\mbox{.}(2022b)]%
        {liu2022welfare}
\bibfield{author}{\bibinfo{person}{Zhihan Liu}, \bibinfo{person}{Miao Lu}, \bibinfo{person}{Zhaoran Wang}, \bibinfo{person}{Michael Jordan}, {and} \bibinfo{person}{Zhuoran Yang}.} \bibinfo{year}{2022}\natexlab{b}.
\newblock \showarticletitle{Welfare maximization in competitive equilibrium: Reinforcement learning for markov exchange economy}. In \bibinfo{booktitle}{\emph{International Conference on Machine Learning}}. PMLR, \bibinfo{pages}{13870--13911}.
\newblock


\bibitem[Mazur(1985)]%
        {mazur1985probability}
\bibfield{author}{\bibinfo{person}{James~E Mazur}.} \bibinfo{year}{1985}\natexlab{}.
\newblock \showarticletitle{Probability and delay of reinforcement as factors in discrete-trial choice}.
\newblock \bibinfo{journal}{\emph{Journal of the Experimental Analysis of Behavior}} \bibinfo{volume}{43}, \bibinfo{number}{3} (\bibinfo{year}{1985}), \bibinfo{pages}{341--351}.
\newblock


\bibitem[Mi et~al\mbox{.}(2023)]%
        {mi2023taxai}
\bibfield{author}{\bibinfo{person}{Qirui Mi}, \bibinfo{person}{Siyu Xia}, \bibinfo{person}{Yan Song}, \bibinfo{person}{Haifeng Zhang}, \bibinfo{person}{Shenghao Zhu}, {and} \bibinfo{person}{Jun Wang}.} \bibinfo{year}{2023}\natexlab{}.
\newblock \showarticletitle{Taxai: A dynamic economic simulator and benchmark for multi-agent reinforcement learning}.
\newblock \bibinfo{journal}{\emph{arXiv preprint arXiv:2309.16307}} (\bibinfo{year}{2023}).
\newblock


\bibitem[{National Academies of Sciences, Engineering, and Medicine}(2019)]%
        {NAS2019}
\bibfield{author}{\bibinfo{person}{{National Academies of Sciences, Engineering, and Medicine}}.} \bibinfo{year}{2019}\natexlab{}.
\newblock \bibinfo{booktitle}{\emph{A Roadmap to Reducing Child Poverty}}.
\newblock \bibinfo{publisher}{The National Academies Press}, \bibinfo{address}{Washington, DC}.
\newblock
\showISBNx{978-0-309-48398-8}
\urldef\tempurl%
\url{https://doi.org/10.17226/25246}
\showDOI{\tempurl}


\bibitem[Samuelson(1937)]%
        {samuelson1937note}
\bibfield{author}{\bibinfo{person}{Paul~A Samuelson}.} \bibinfo{year}{1937}\natexlab{}.
\newblock \showarticletitle{A note on measurement of utility}.
\newblock \bibinfo{journal}{\emph{The review of economic studies}} \bibinfo{volume}{4}, \bibinfo{number}{2} (\bibinfo{year}{1937}), \bibinfo{pages}{155--161}.
\newblock


\bibitem[Strotz(1973)]%
        {strotz1973myopia}
\bibfield{author}{\bibinfo{person}{Robert~Henry Strotz}.} \bibinfo{year}{1973}\natexlab{}.
\newblock \bibinfo{booktitle}{\emph{Myopia and inconsistency in dynamic utility maximization}}.
\newblock \bibinfo{publisher}{Springer}.
\newblock


\bibitem[Svensson(2020)]%
        {svensson2020monetary}
\bibfield{author}{\bibinfo{person}{Lars~EO Svensson}.} \bibinfo{year}{2020}\natexlab{}.
\newblock \bibinfo{booktitle}{\emph{Monetary policy strategies for the Federal Reserve}}.
\newblock \bibinfo{type}{{T}echnical {R}eport}. \bibinfo{institution}{National Bureau of Economic Research}.
\newblock


\bibitem[Tesfatsion and Judd(2006)]%
        {tesfatsion2006handbook}
\bibfield{author}{\bibinfo{person}{Leigh Tesfatsion} {and} \bibinfo{person}{Kenneth~L Judd}.} \bibinfo{year}{2006}\natexlab{}.
\newblock \bibinfo{booktitle}{\emph{Handbook of computational economics: agent-based computational economics}}.
\newblock \bibinfo{publisher}{Elsevier}.
\newblock


\bibitem[{U.S. Bureau of Labor Statistics}(2023)]%
        {bls}
\bibfield{author}{\bibinfo{person}{{U.S. Bureau of Labor Statistics}}.} \bibinfo{year}{2023}\natexlab{}.
\newblock \bibinfo{title}{Employment Situation}.
\newblock \bibinfo{howpublished}{\url{https://www.bls.gov/news.release/empsit.toc.htm}}.
\newblock


\bibitem[Wheat et~al\mbox{.}(2022)]%
        {JPMC}
\bibfield{author}{\bibinfo{person}{Chris Wheat}, \bibinfo{person}{Erica Deadman}, {and} \bibinfo{person}{Daniel~M Sullivan}.} \bibinfo{year}{2022}\natexlab{}.
\newblock \bibinfo{booktitle}{\emph{How families used the advanced Child Tax Credit.}}
\newblock \bibinfo{type}{{T}echnical {R}eport}. \bibinfo{institution}{JPMorgan Chase Institute}.
\newblock


\bibitem[Zheng et~al\mbox{.}(2022)]%
        {zheng2022}
\bibfield{author}{\bibinfo{person}{Stephan Zheng}, \bibinfo{person}{Alexander Trott}, \bibinfo{person}{Sunil Srinivasa}, \bibinfo{person}{David~C. Parkes}, {and} \bibinfo{person}{Richard Socher}.} \bibinfo{year}{2022}\natexlab{}.
\newblock \showarticletitle{The {AI} Economist: {T}axation policy design via two-level deep multiagent reinforcement learning}.
\newblock \bibinfo{journal}{\emph{Science Advances}} \bibinfo{volume}{8}, \bibinfo{number}{18} (\bibinfo{year}{2022}), \bibinfo{pages}{2607}.
\newblock
\urldef\tempurl%
\url{https://doi.org/10.1126/sciadv.abk2607}
\showDOI{\tempurl}


\end{thebibliography}

\end{document}